\documentclass[11pt]{article}
\usepackage[left=1in,right=1in,top=1.2in,bottom=1.2in,
            footskip=.25in]{geometry}
\usepackage{amsmath, amssymb,amstext} 
\usepackage{url}
\usepackage{amsmath}
\usepackage[table]{xcolor}
\usepackage{authblk}
\usepackage{amssymb}
\usepackage{subfigure}
\usepackage{ragged2e}
\usepackage{hyperref}
\usepackage{graphics}
\usepackage{mathtools}
\usepackage{caption}
\usepackage{subcaption}
\usepackage{algpseudocode}
\usepackage{algorithm}

\usepackage{afterpage}
\usepackage{makecell}
\usepackage{natbib}
\usepackage{array}
\usepackage{collcell}
\usepackage{tcolorbox}
\usepackage{multirow}
\title{Dispersion based Recurrent Neural Network Model for Methane Monitoring in Albertan Tailings Ponds} 

\author[1,a]{Esha Saha}
\author[3]{Oscar Wang}
\author[1]{Amit K. Chankraborty}
\author[1]{Pablo Venegas Garcia}
\author[1]{Russell Milne}
\author[1,a]{Hao Wang}

\affil[1]{Interdisciplinary Lab for Mathematical Ecology and Epidemiology (ILMEE) \& The Department of Mathematical and Statistical Sciences, University of Alberta, Edmonton (AB), T6G 2J5, Canada}
\affil[3]{The Department of Mathematics and Computer Science, University of Richmond, Richmond (VA), 23173, United States}
\affil[a]{Corresponding authors. E-mail: esaha1@ualberta.ca, hao8@ualberta.ca}
\date{}

\begin{document}
\maketitle
\begin{abstract}
Bitumen extraction for the production of synthetic crude oil in Canada's Athabasca Oil Sands industry has recently come under spotlight for being a significant source of greenhouse gas emissions. 
A major cause of concern is methane, a greenhouse gas produced by the anaerobic biodegradation of hydrocarbon in oil sands residues, or tailing, stored in settle basins commonly known as oil sands tailing ponds.
In this work, we build a data-driven modeling framework to determine the methane emitting potential of these tailing ponds and have future methane projections using a Dispersion based Recurrent Neural Network (DIRNN).
We show that our method can predict both methane emissions and concentrations by considering the \textcolor{black}{transport of methane emissions in air}, thereby outperforming existing other deep learning approaches. Using a reverse dispersion modeling approach, we use our trained model to identify active ponds and estimate \textcolor{black}{about 56,303 tonnes of methane (1.5 million tonnes of carbon dioxide equivalent) emissions from the Athabasca oil sands tailings. Our results are consistent with previously reported emission estimates from various studies, and indicate atleast three times underestimation in official reports. } 

\end{abstract}
\section{Introduction}
Anthropogenic sources of greenhouse gases (GHGs) are the major drivers of climate change with methane (CH$_4$) having the second largest share of emissions in the atmosphere after carbon dioxide (CO$_2$). Although the comparative impact of CH$_4$ is 28 times greater than CO$_2$ over a 100-year period \citep{usepa}, it has a shorter lifespan of 12 years, which makes CH$_4$ mitigation policies a cost-effective short-term approach to combat global warming \citep{Flannigan2009,airqualityshapleypaper}. 
 Oil sands activities contribute significantly to GHG emissions and in particular are regarded as sources of pollution \citep{schindler2014unravelling,liggio2016oil,yu2023tailings}. 
 The mining and extraction of oil sands is directly associated with deforestation and release of sulfur oxides, nitrogen oxides, hydrocarbons, and fine particulate matter, etc.
The oil sand tailing ponds (OSTPs) contain toxic industrial wastes that can leak into fresh water sources affecting aquatic ecosystems. Moreover, OSTPs emit significant quantities of CH$_4$ from toxin degradation by anaerobic bacteria \citep{siddique2007metabolism,michel2024rapid} leading to elevated levels of CH$_4$ in the air. 
Frequent onsite data collection for a detailed understanding of OSTPs is often infeasible due to expensive measurement technology, toxic air quality or hard to obtain permit requirements \citep{airqualityshapleypaper}.

Prior OSTP related methane modeling efforts involve controlled laboratory experiments, whose results are used to build mechanistic models (MM) of CH$_4$ emissions (total mass of CH$_4$ released in atmosphere from source in a time interval) from toxin/hydrocarbon degradation by bacteria in OSTPs \citep{kong2019second,pablo}. 
Since experimental limitations lead to exclusion of many relevant parameters such as temperature, pressure, wind speed, etc., MMs cannot be used to accurately assess methane concentration levels (the amount of CH$_4$ in the air at a given place and time). Infact, the relationship between emissions and concentrations can be better described by atmospheric dispersion models (ADMs) \citep{stockie2011mathematics,das1998release,mohandevelopment}. 
However, simple ADMs like the Gaussian plume or puff models \citep{stockie2011mathematics,mikkelsen1987diffusion} are unrealistic due to multiple modeling assumptions, while the realistic formulations are highly complex and non-linear, and may require high computational resources to be solved numerically.
For example, refer to the model formulation of AERMOD \citep{cimorelli2005aermod}, a well-known software developed by Environmental Protection Agency (EPA). One needs an understanding and data availability of multiple parameters woven through 30+ equations in order to solve the model. Other similar softwares include CALPUFF, FLEXPART, WindTrax, etc \citep{bakels2024flexpart,bonifacio2013comparison,tagliaferri2022validation}.

\textcolor{black}{Data-driven techniques, on the other hand, offer a cost-effective and powerful alternative to classical modeling in the field of Environmental Sciences, both from the perspective of modeling complex phenomenon as well from the socio-economic modeling of policy gains.}
\textcolor{black}{From a socio-economic and policy standpoint, data-driven approaches can enable low-cost, high-resolution tracking of industrial methane emissions, essential for stakeholders such as regulators, environmental agencies, and oil sands operators. 
Moreover, machine learning techniques are increasingly being applied to stock markets and energy pricing models \citep{kumar2025comparison,basher2025important,jin2024price,cheng2025novel} including coal and carbon pricing models.
These models help policymakers and analysts simulate the effects of emissions-based taxation under various economic conditions by capturing the non-linear dependencies between fuel demand, carbon markets, and policy shifts \citep{islam2025role,jin2,d2025evaluating,jin2025bayesian}.
These applications not only inform government strategy but also assist energy firms in optimizing operations.}

\textcolor{black}{For applications in modeling environmental phenomenon, some well-known machine learning architectures used for analyzing and predicting various atmospheric gases include Random Forests (RFs), neural networks(NN), recurrent neural networks (RNNs), long short-term memory (LSTM), bidirectional LSTM, stacked
LSTM, and gated recurrent unit (GRU) \citep{xie2019research,tong2019deep,luo2023machine,hu2021estimating,hamrani2020machine,meng2022methane,hou2022revealing}. Other data-driven methods use satellite images as an effective way to build a dataset for the purpose of training classifiers to identify active OSTPs so that only high risk ponds may be closely monitored \citep{yu2023tailings,psomouli2023methane,qu2024inverse}.
Some recent works have used used satellite data to detect CH$_4$ emissions across Canada \citep{yazdinejad2025advanced}. However, data collected by satellites is expensive, hard to process and sparse making them less useful for models where frequent data collection is the key to proper model training.
Machine learning models can also be trained for effective risk prediction of OSTPs as an alternative to standard monitoring systems, which are often expensive and have poor lightning protection abilities \citep{yang2020effective}. 
While most existing models (both mechanistic and data-driven) can perform well on estimating emissions from tailing ponds or predicting concentrations, an understanding of how atmospheric concentrations of CH$_4$ gets affected by emission sources are limited \citep{koushafar2023deep}.} 

\textcolor{black}{ 
 In order to fill this gap and track emission sources, we propose a novel Dispersion based Recurrent Neural Network (DIRNN) framework that can successfully use the dynamics of CH$_4$ emissions from OSTPs to predict concentrations in the atmosphere by preserving the physics of CH$_4$ dispersion in air.}
\textcolor{black}{With an unique hybrid approach yet to be used widely in environmental modeling, given a dataset consisting of atmospheric variables (collected from weather monitoring stations around an OSTP) and hydrocarbon degradation (simulated from solving MMs) in an OSTP, our dispersion informed trained model can successfully predict CH$_4$ emissions and concentration levels near the OSTPs, outperforming other deep learning models.
Instead of treating these two quantities separately, the proposed model enforces constraints based on atmospheric dispersion (advection‑diffusion dynamics) to tie together emission and concentration. 
Model training using constrained optimization with penalty ensures balance between data fitting and physical consistency.
Further, we show that our trained model is capable of giving emission estimates for various OSTPs, irrespective of whether they were included in the training dataset or not, as long as they are in a close proximity to each other. 
This is made possible due to the reverse dispersion penalty from physical equations that enables daily emission estimation from multiple directions around weather stations, without direct sampling of each site.
The novelty of our proposed model lies in various aspect of the framework which include (but are not limited to) model design (physics-constrained neural network using atmospheric transport), emission inference (inverse modeling of CH$_4$ sources from ambient data and wind direction), source attribution (directional decomposition of emissions), joint modeling
(emissions and concentrations estimated simultaneously) and discovery of underestimated emissions, including from inactive ponds.}

\textcolor{black}{
The paper is arranged as follows. We discuss the region of interest, data collection techniqies and processing in Section \ref{sec:study-area}, followed by the model framework in Section \ref{sec:methods}. The results are discussed in Section \ref{sec:results} with the discussion and conclusions in Sections \ref{sec:discussion} and \ref{sec:conclusion}, respectively.}

\section{Study Area, Data Collection and Preprocessing}\label{sec:study-area}
\textcolor{black}{Oil sands tailings are the by-products generated after separating bitumen from oil
sands. These tailings are stored in large engineered reservoirs known as oil sands
tailings ponds (OSTPs) \citep{ostp1}. They are composed of a mixture of sand, silt, clay fines,
process water, and small amounts of unrecovered hydrocarbons from the extraction
process \citep{ostp2}. When first deposited into an OSTP, fresh tailings are mostly water ($\sim$ 85\%),
containing about 8 \% mineral fines and less than 1\% unrecovered hydrocarbons.
Over time, the fine particles gradually consolidate, eventually forming dense layers
called mature fine tailings (MFT) with more than 30 \% solids near the pond bottom \citep{ostp1}. 
Tailings
temperature varies with depth, ranging from roughly 12$^\circ$C at 6 meters (m) to about 22$^\circ$C at 30
m below the mudline \citep{ostp3,ostp4}. 
The process water is typically alkaline, with a pH of about 8.5. 
Chemically, tailings contain residual hydrocarbons and
soluble electron acceptors such as sulfate and iron; in some cases, gypsum is added to accelerate consolidation. 
Direct data collection from OSTPs is challenging due to hard to obtain permit requirements.}

\textcolor{black}{Given a dataset consisting of atmospheric variables (collected from weather monitoring stations around an OSTP) and hydrocarbon degradation (simulated from solving the MMs built using laboratory controlled experiments) in an OSTP, we are interested in building a machine learning model that can predict both methane concentration and emission simultaneously.
Thus, the input to the proposed framework includes various parameters that directly or indirectly affect atmospheric methane concentrations. 
The model considers three types of input data: (i) $\mathbf{x}_{dil}$ denoting the degradation of hydrocarbons in OSTPs and obtained from solving MMs in literature; (ii) $\mathbf{x}_{atm}$ representing atmospheric parameters such as ambient temperature, wind speed, wind direction, solar activity, etc; and (iii) time vector $\mathbf{t}$. 
These three inputs $\mathbf{x}_{dil}$, $\mathbf{x}_{atm}$, and $\mathbf{t}$ together form the input $\mathbf{x}$ and are used to define the model and its corresponding minimization problem.}
In this section, we discuss the study area (Section \ref{sec:study}) and the technique of dataset building.
The meteorological data collection from weather monitoring stations is discussed in Section \ref{sec:wmd} and simulated data from experimentally validated MM (which estimates methane emissions and hydrocarbon degradation in active  OSTPs) is discussed in Section \ref{sec:mm}.

\subsection{Study Area}\label{sec:study}
\begin{figure}[h!]
    \centering
    \includegraphics[scale=0.4]{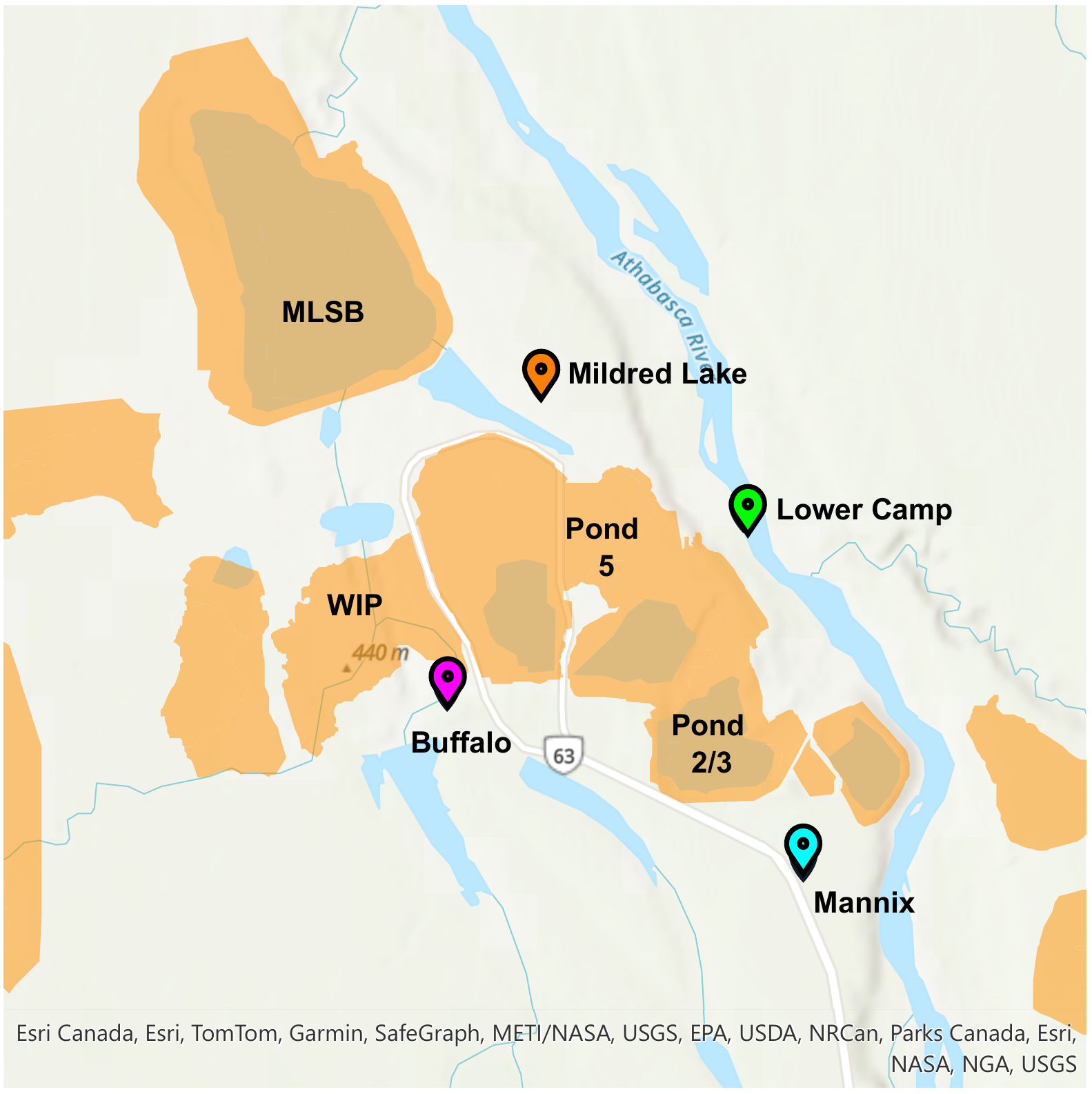}
    \caption{Region of Wood Buffalo with all the weather monitoring stations of interest and main OSTPs and/or EPLs. \textcolor{black}{Selected stations: Mannix (`Pond 2/3' approximately 1.4 km northwest of the station); Lower Camp (`Pond 2/3' approximately 3.5 kms southwest and `Pond 5' about 1.4 kms west of the station, respectively); Mildred Lake (`Mildred Lake Settling Basin'/MLSB approximately 1.7 kms northwest and `Pond 5' approximately 2.1 kms southeast of the station, respectively); Buffalo (`West-In-Pit'/WIP at 0.8 kms northwest of the station). }}
    \label{fig:study_area}
\end{figure}

Our region of interest is located in the industrial area around Syncrude and Suncor Base Plants in the Athabasca Oil Sands deposits. The region contains multiple weather monitoring stations (some located near the oil sands mining areas) under the Wood Buffalo Environmental Association (WBEA) \citep{wbeaDatabase} that measures hourly data of the ambient air quality and meteorological parameters. 
Overall elevation of the WBEA region is about 200-300 m above sealevel.
Lower Camp is located by the Athabasca River Valley at about 115 m south
of the Syncrude pump house and 238 m above sealevel. It has an active OSTP `Pond 2/3' approximately at a distance of 3.5 kms southwest of the station and an abandoned OSTP `Pond 5' approximately at 1.4 kms, both owned by Suncor.
Located at 332 m above sea level, Mannix station is less than 5 km from the Suncor base plant whose land use segregation is between 0-180 degrees of the station. There are no airflow restrictions and an active OSTP `Pond 2/3' owned by Suncor is located approximately 1.4 kms northwest of the station.
Mildred Lake is Located within 400 m of the Syncrude airstrip, the station sits at 314 m above sea level and within 5 km from the Syncrude base plant on West. There is partial restriction of airflow in North, South and West of the station by buildings and/or trees which lie within 40-160 m. It has been measuring methane from December 2019 onwards. An active OSTP `Mildred Lake Settling Basin'/MLSB is approximately 1.7 kms northwest of the station, owned by Syncrude, and inactive `Pond 5' is approximately 2.1 kms owned by Suncor.
Buffalo station sits at 315 m above sea level and less than 5 km from the Syncrude base plant and 0.8 km from an OSTP. There is no restriction of airflow. The land use segregation reports oil sands plant in 0-90 degrees and 271-360 degrees of the station. It's nearest pond, `West-In-Pit'/WIP is presently converted to an EPL and sits at a distance of 0.8 kms northwest of the station. 
The study area is given in Figure \ref{fig:study_area}.

Note that out of the four selected stations, we build datasets using the stations that are associated with `active' OSTPs only, because source estimations using MMs are based on experiments modeled after `active' OSTPs. We classify an OSTP to be `active' if it there is continuous inflow of diluents (industrial residues). 
Based on available data and information on OSTPs, we select stations Mannix with Pond 2/3, Lower Camp with Pond 2/3 and Mildred Lake with MLSB for building the dataset.

\subsection{Dataset Creation from On-Field Data}\label{sec:wmd}

For $d>1$ and number of days $k$, suppose the input-output pairs are denoted by $\{(\mathbf{x}_n,\mathbf{y}_n)\}_{n=1}^k$, with $\mathbf{x} = [\mathbf{x}_{atm},\mathbf{x}_{dil}]\in\mathbb{R}^d$ and $\mathbf{y} = [y^{conc},y^{emm}]\in\mathbb{R}^2$. The vector $\mathbf{x}_n$ consists of various input variables sampled across $k$ timesteps (days) and the two-dimensional vector $\mathbf{y}_n$ consists of the CH$_4$ emissions, and concentrations. 
For each $k$, the input vector $\mathbf{x}_n$ is built using data from two sources: atmospheric variables $\mathbf{x}_{atm}$ and industrial/chemical component variables $\mathbf{x}_{dil}$. 
Dataset for atmospheric parameters $\mathbf{x}_{atm}$ and methane concentrations $y_n^{conc}$ is built from data collected by the Wood Buffalo Environmental Association (WBEA) \citep{wbeaDatabase}.
The training dataset is built only for stations near an active OSTP (MLSB and Pond 2/3 as reported in \citep{Burkus2014}) (as the mechanistic models can only mimic the kinetics of an active tailings pond). 
We pick the stations in a close proximity to these active OSTPs that have no other methane sources (for example, wetlands) between them. 

For each station, we build the dataset by filtering observations with wind direction ranges based on location of active OSTPs i.e., Mannix (300-320 degrees), Lower Camp (160-180 degrees), and Mildred Lake (300-340 degrees). 
The data from atmospheric variables denoted by $\mathbf{x}_{atm}$ includes temperature, pressure, relative humidity, solar radiation, etc and is collected hourly by weather monitoring stations under consideration.
Similarly, the WBEA dataset reporting CH$_4$ concentrations is used to build one of the output dataset samples, $y_n^{conc}$.
The map of the weather stations and OSTPs is given in Figure \ref{fig:study_area}. All variables used in the input $\mathbf{x}_{atm}$ is summarized in Table \ref{tab:var}.

\subsection{Data from Experimentally Validated Model(s)}\label{sec:mm}
To build dataset for $\mathbf{x}_{dil}$ and source emissions $y_n^{emm}$, we solve MMs describing the properties and dynamics of methane production in OSTPs, with appropriate parameters and initial conditions \citep{kong2019second,firstorder,pablo}. 
These experimentally validated models are developed by considering the most labile hydrocarbons present in the diluents/solvents used by each of the oil sands companies. 
The models are generally represented by a dynamical system whose general form is described in Eq. \eqref{eq:mm}. 
The system describes the degradation dynamics of each of the labile hydrocarbon by methanogenic bacteria. 
For each fixed $i$ (the value of $i$ depends on the number of labile hydrocarbons considered), the system of equations are given as
\begin{eqnarray}\label{eq:mm}
    \dfrac{dC_i}{dt} &= & f(C_i,t,y_1,\cdots,y_k) \nonumber\\
    \dfrac{dy_j}{dt} &= & g_j(C_i,t,y_j) \text{ for } j=1,\cdots,k\\
    \text{CH}_4 &= & h(C_i,y_1,\cdots,y_k,\mu_i,t) \nonumber
\end{eqnarray}
where $C_i$ denotes each of the hydrocarbons, $y_j$ denotes other variables in consideration (for example, other nutrients, biomass of microbes, etc) and $\mu_i$ denotes the set of constants corresponding to methane production (for example, microbial efficacy, stoichiometric factor, etc).
The exact nature of the functions $f,g$ and $h$ along with the other parameters can be found in prior works on modelling methane emissions from OSTPs \citep{kong2019second,firstorder,pablo}.

In order to solve the MM, we use the monthly `Flared and Wasted' category of ``diluents" from \citep{st39} as the total monthly inflow of hydrocarbons into the ponds. 
Based on prior works that have studied possible chemical composition of these diluents \citep{kong2019second,siddique2006biodegradation}, we split this diluent data into about 20 labile hydrocarbons.
Note that we consider only a small fraction of the diluents based on the OSTP Fine Fluid Tailings (FFT) volume (20\% for MLSB and 15\% for Pond 2/3 i.e., only this much amount of total reported diluents is assumed to be tossed into the OSTPs) and divide it by the number of days per month to estimate daily inflow of diluents in the MM. 
These hydrocarbons are assumed to be subsequently used up by methanogenic bacteria leading to methane emissions.  
 A constant daily inflow of diluents is assumed i.e. the total monthly diluent reported in \citep{st39} is divided equally by the number of days in the month (to obtain the initial conditions). The system is solved with a timestep of one day for a month. This is done for all the months from January 2020 to December 2023 in order to build the dataset.
 This approach is similar to the technique followed in \citep{pablo}.
 The hydrocarbon degradation described by the MMs forms the dataset $\mathbf{x}_{dil}$.
We experiment with $\mathbf{x}_{dil}$ built based on two methanogenesis models \citep{pablo,kong2019second}.
The CH$_4$ values obtained using these models form $y^{emm}$.
Note that when generating the data using the model proposed in \citep{kong2019second}, we identified a small number of generated daily values (about 1\% of values) that were not biologically realistic. We considered these values to be numerical artifacts of simulation driven by the stiffness of the model, and hence replaced each one with the generated value for the day before it for the sake of simplicity and biological fidelity.

Once all types of data is obtained, they are averaged and interpolated to obtain sample points at the frequency of one data per day. For example, variables from $\mathbf{x}_{atm}$ are collected hourly, so they are averaged over 24 hours to obtain one sample per day. On the other hand, diluent data is reported monthly by companies and hence interpolation technique is used to fit a spline function and sample daily data points.
 
\section{Problem Formulation and Framework} 
\label{sec:methods}
\begin{figure}
    \centering
    \includegraphics[width=0.3\linewidth]{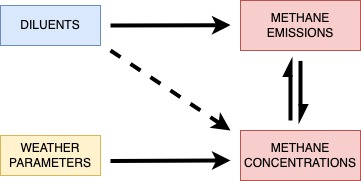}\,\includegraphics[width=0.7\linewidth]{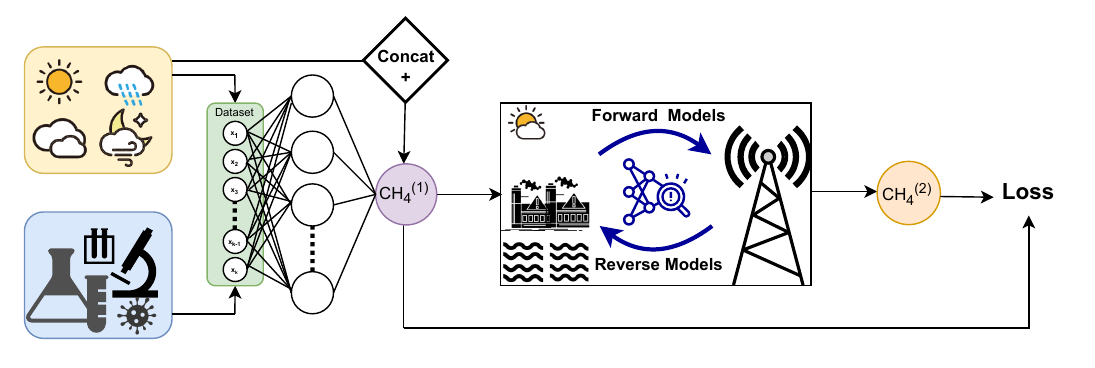}
    \caption{Graphical representation of the proposed modeling framework. \textbf{Left: }Representation of how the different input variables interact. Solid lines depict a direct connection and the dashed lines represent an indirect affect. \textbf{Right:} A flowchart of DIRNN.}
    \label{fig:flow}
\end{figure}

The goal of the proposed research is to train a parameterized model to track emissions from OSTPs using CH$_4$ concentrations.
\textcolor{black}{
We incorporate the dynamics from MMs of OSTPs and physical constraints from atmospheric dispersion models to train a machine learning framework over a give period of time. 
The idea is to model the interactions depicted in Figure \ref{fig:flow}.
\begin{itemize}
    \item Diluents directly affect emissions from OSTPs. For example, more diluents would directly lead to more hydrocarbon degradation and thus more CH$_4$ emissions.
    \item CH$_4$ emissions and concentrations are directly related to each other. Increase in one of the quantities will automatically lead to an increase in the other and vice versa.
    \item Atmospheric parameters directly affect the concentration of CH$_4$ but do not significantly affect the formation of CH$_4$ in the OSTPs (for example, the wind direction or speed has nothing to do with how the microbes degrade hydrocarbons at the bottom layer of the OSTPs).
    \item Diluents indirectly affect the concentrations by increasing emissions which in turn increases the concentrations.
\end{itemize}}

Thus, we combine all the variables (diluents and atmospheric data) and use it as an input to a machine learning model. The output is then combined with the data from atmospheric variables (only) and used as an input to another machine learning model which is defined using the atmospheric dispersion model equation(s). A detailed flowchart is given in Figure \ref{fig:flow}.

To optimize the weights and biases of DIRNN, the dispersion constrained optimization problem \citep{xu2022physics,saha2023spade4,kashinath2021physics,antonion2024machine} is modified to include information from MMs as well as real-time data affecting concentrations. 
For each $i$th observations in the set $\mathcal{I}_{obs}$, suppose  $\mathbf{x}_i$ denotes the $(d+1)-$dimensional input vector and $u_i$ denotes observed output. Then given a fixed function $q:\mathbb{R}^{d+1}\rightarrow\mathbb{R}$ describing emission dynamics (from MMs), our modified constrained optimization problem aims to find a function $u:\mathbb{R}^{d+1}\rightarrow \mathbb{R}$ by solving the problem,
\begin{equation}\label{eq:1}
    \min\limits_{\phi} \frac{1}{|\mathcal{I}_{obs}|}\sum\limits_{i\in\mathcal{I}_{obs}}(u(\mathbf{x}_i) - u_i)^2 \text{ subject to }F(\phi(\mathbf{x},u),u,q)=0,
\end{equation}
 where $F$ is the physical constraint with unknown function $\phi$. Since we want to learn $u$ from given measurements and $\phi$ is unknown, we parameterize them with $u_{\bar{\Theta}}(\mathbf{x})$ and $\phi_{\hat{\Theta}}(\mathbf{x},u)$, respectively where $\hat{\Theta}$ and $\bar{\Theta}$ denote the unknown parameters to be learned through training.
Different function representation models such as sparse polynomial approximation \citep{schaeffer2018extracting}, random feature models \citep{saha2023harfe}, or neural networks \citep{mcculloch1943logical} may be used. Converting Eq. \eqref{eq:1} to an unconstrained optimization problem we get,
\begin{equation}\label{eq:optimization_original}
    \min\limits_{\bar{\Theta},\hat{\Theta}} \frac{1}{|\mathcal{I}_{obs}|}\sum\limits_{i\in\mathcal{I}_{obs}}\left[(u_{\bar{\Theta}}(\mathbf{x}_i) - u_i)^2 + \lambda\left(F(\phi_{\hat{\Theta}}(\mathbf{x}_i,u_{\bar{\Theta}}),u_{\bar{\Theta}},q(\mathbf{x}_i))\right)^2\right],
\end{equation}
where $\lambda\in (0,\infty)$. An advantage of using this method (also known as penalty method) is that it avoids solving the constraint $F(\phi, u,q) = 0$ \citep{xu2022physics}. However, the physical constraints may not be satisfied exactly i.e., theoretically, $F(\phi, u,q) = 0$ will hold only when $\lambda\rightarrow \infty$ \citep{xu2022physics}. 
 While optimizing the choice of $\lambda$, it is important to remember that a large value of $\lambda$ places less weight on the objective function. 
 Hence, a proper choice of $\lambda$ is based on the desirable trade-off between fitting the observed value and satisfying the constraint. This technique a has been adapted in numerous works involving learning of systems of PDEs/ODEs from data \citep{raissi2019physics}.
Our choice of the constraint $F$ is derived from an atmospheric dispersion models called Gaussian Plume Model (GPM) \citep{stockie2011mathematics}, 
\[\dfrac{\partial u}{\partial t} + \nabla.J = q,\]
where $u(\vec{x},t)$ is the mass concentration, $q(\vec{x},t)$ is a source (or sink) and $J$ is mass
flux due to diffusion ($J_D$) and advection ($J_A$), and $\vec{x}$ and $t$ denote space and time respectively.
Assuming negligible sinks, the function $q$ defined from the MMs acts as the source term. 
Since we fix the coordinates of the source (OSTPs) and weather stations in space, $u$ and $q$ are independent of spatial coordinates.
Thus, we are interested in exploring the relationship of these two functions ($u$ and $q$) to various input variables involving atmospheric parameters and/or hydrocarbon degradation.

Given the input dataset $\mathbf{x} =[\mathbf{x}_{atm},\mathbf{x}_{dil},\mathbf{t}]^T\in\mathbb{R}^{d+1}$, where $\mathbf{x}_{dil}$ is built from simulated data from MMs and $\mathbf{x}_{atm}$ is built from the real-time weather station measurements, the CH$_4$ concentrations and emissions are trained using the parameterization below in Eq. \eqref{eq:DIRNN_for1} and \eqref{eq:DIRNN-for}. 
Output of the first network $u_{\bar{\Theta}}$ gives the predicted methane concentration. 
We then concatenate this output with atmospheric parameters $\mathbf{x}_{atm}$ (since those are the only variables that can potentially affect diffusion/advection), and use it as an input to the second neural network that estimates $\nabla.J$. 
The exact form of the outputs are given by
\begin{eqnarray}
    u_{\bar{\Theta}} &=& \bar{\Theta}_4 \sigma\left(\bar{\Theta}_3\sigma\left(\bar{\Theta}_2\sigma\left(\bar{\Theta}_1\mathbf{x}\right)\right)\right)\label{eq:DIRNN_for1}\\
q_{\theta}(\mathbf{x},t) &=& \text{Grad}_t\left(u_{\bar{\Theta}}\right) + \hat{\Theta}_3\sigma\left(\hat{\Theta}_2\sigma\left(\hat{\Theta}_1\left[
        u_{\bar{\Theta}} ,
        \mathbf{x}_{atm}]
    \right]\right)\right),\label{eq:DIRNN-for}
\end{eqnarray}
where $\sigma$ denotes the activation function, Grad$_t(\cdot)$ denotes the partial time derivative of the output $u_{\bar{\Theta}}$ and $q_{\theta}(\mathbf{x},t)$ denotes the learned parameterized source emissions. Alternatively, the function to define $\nabla.J$ can also be represented with a known basis such as a polynomial basis. This can be used when trying to decipher the important variables affecting the diffusion/advection terms. Eq. \eqref{eq:DIRNN-for} thus modifies to 
\[ q_{\theta}(\mathbf{x},t) = \text{Grad}_t\left(u_{\bar{\Theta}}\right) + \hat{\Theta}_1 \mathcal{P}\left(\left[
        u_{\bar{\Theta}} ,
        \mathbf{x}_{atm}]
    \right]\right),\]
where $\mathcal{P}([z_1,\cdots,z_n]) = [1,z_1,\cdots,z_n,z_1z_2,\cdots,z_{n-1}z_n,z_1^2,\cdots,z_n^2]$.

\subsection{Inverse Dispersion based Framework (iDIRNN) for Emission Estimates}
In order to estimate source emissions and identify active tailing, we modify the constraint to satisfy inverse dispersion models that aim to quantify emissions when concentrations are given. Given a source emitting gas at a continuous and unknown rate $q$ $kg/s$, suppose the time-average gas concentration above background denoted by $C - C_b$ (where $C$ is the measured concentration and $C_b$ is the background concentration) is measured at some point $M$. Then the emission using inverse dispersion modeling is given as

\[q = (C/Q)_{sim}^{-1} (C-C_b),\]

where $(C/Q)_{sim}$ denotes the ratio of concentration at $M$ to the source emission rate predicted by an atmospheric dispersion model. While the equation seems pretty straightforward, prediction of $(C/Q)_{sim}$ is not trivial and often ill-conditioned. Different types of dispersion models (e.g., Gaussian plume, K-theory) make this calculation with different levels of sophistication. 
Realistically, models should take into account average wind and turbulence statistics of the atmosphere along with possible dispersion of the source and how it relates to the concentrations. Various numerical methods as well as prior emission estimates may be needed to solve the problem \citep{vojta2022comprehensive}.

In iDIRNN, using inputs $[\mathbf{x}_{dil},\mathbf{x}_{atm},\mathbf{t}]^T\in\mathbb{R}^{d+1}$ we still learn $u_{\bar{\Theta}}$ from fitting the model to observed concentrations as in Eq. \eqref{eq:DIRNN_for1}, however the constraint is now based on finding source emissions $q$ from the above equation.
We parameterize the inverse of influence function  $(C/Q)_{sim}$ as a neural network so that the constraint in Eq. \eqref{eq:DIRNN-for} becomes,
\begin{eqnarray}\label{eq:DIRNN_rev}
   q_{\theta}(\mathbf{x},t) = \hat{\Theta}_3\sigma\left(\hat{\Theta}_2\sigma\left(\hat{\Theta}_1\left[
        u_{\bar{\Theta}} ,
        \mathbf{x}_{atm}
    \right]\right)\right).
\end{eqnarray}
Note that in this formulation, the term $C_{b}$ term is balanced out by the bias terms present in estimating emissions and concentrations. 
This offers multiple advantages: (1) the formulation lets the model automatically learn from given data; (2) it is useful for various applications where data and understanding of the dynamics is limited.
We first train the model on our datasets. In order to get emission estimates, we replace $u_{\bar{\Theta}}$ in Eq. \eqref{eq:DIRNN_for1} with the concentration data measured by the weather station into the trained model. 
The emission estimates per day are then added up to get cumulative emissions over a given year. 

\subsection{Model Architecture, Training and Validation Analysis}
\begin{algorithm}[ht!]
\caption{Model training using proposed framework}\label{alg:code}
\begin{algorithmic}[1]
\Require Input data $\mathbf{x} = [\mathbf{x}_{dil},\mathbf{x}_{atm},\mathbf{t}]$, observed data $u$, emission function $q(\mathbf{x})$, observations indices set $\mathcal{I}_{obs}$, physical constraint $F$, model architectures for learning $u_{\bar{\Theta}}$ and $\phi_{\bar{\Theta}}$, penalty parameter $\lambda$.
\Statex{}
\For {$j$ in \texttt{epochs}}
\State{$y = u_{\bar{\Theta}}(\mathbf{x})$; $z = \phi_{\bar{\Theta}}(y,\mathbf{x}_{atm})$}.
\State Update $\Theta = [\bar{\Theta},\hat{\Theta}]$ by minimizing the loss,
\begin{equation*}
    \frac{1}{|\mathcal{I}_{obs}|}\|y-u\|_2^2+ \frac{\lambda}{|\mathcal{I}_{obs}|}\left\|F(z,y,q)\right\|_2^2
 \end{equation*}
 \If {Sparse Parameters == True}
 \For {k in size($\hat{\Theta}$)}
 \If {$|\hat{\Theta}_k|<10^{-4}$}
 \State{$\hat{\Theta}_k=0$}
 \EndIf
\EndFor
 \EndIf
\EndFor
\Statex{\textbf{Output:} Concentrations $y$ and emissions $z$}
\end{algorithmic}
\end{algorithm}
The entire dataset contains 1096 daily measurements dated between January 1, 2020 to December 31, 2022. Dataset for the year 2023 is built separately for testing the trained model(s), consisting of 365 samples to predict emissions and concentrations from the trained model. 
The training dataset is standardized between 0 and 1 to avoid unnecessary bias of input features with a larger scale. 
The conversion between different units of measurement between emission and concentration data are implicitly taken care of while model training. 
The model was trained on 80\% of the samples from the dataset with upto 10,000 iterations.
The training loss for all the stations fell below $10^{-4}$. 
Given the size of our dataset, we use three layers with 500 and 200 neurons each to learn $u_{\bar{\Theta}}$ and $\phi_{\hat{\Theta}}$, respectively.
For optimizing the weights, we use a stochastic gradient descent (SGD) algorithm with learning rate optimized between $10^{-2}$ and $10^{-3}$, and momentum 0.9. Since the data from weather monitoring stations are noisy, we use an $\ell_2$ based weight decay parameter of $10^{-3}$ in the SGD algorithm to avoid overfitting. 
For 10 random initializations, we train the iDIRNN model and calculate the estimated emissions using real concentration data for the years 2020 and 2021 (true emission data from companies is only available for these two years).
The model whose cumulative CH$_4$ estimations are closest to the true emissions as reported in official documents is picked for further analysis. All codes were written in Python 3.10.9 and can be found at \url{https://github.com/esha-saha/champ}. The pseudocode for model training is given in Algorithm \ref{alg:code}.

We extrapolate our results to obtain total emissions from all OSTPs using the sample mean of emissions computed for the selected OSTPs. 
Since official statistical data reports CH$_4$ emissions in terms CO$_2$ equivalent, we convert our estimations to similar units accordingly for comparison.

\section{Results}\label{sec:results}
We discuss the model simulation outcomes with respect to two objectives: (i) forecasting methane emissions and concentrations from tailing ponds jointly; (ii) identifying active OSTPs and their emissions levels to analyze their impacts with respect to the overall CH$_4$ emissions in Canada.

\subsection{Methane Forecasts from Tailing Ponds}
\begin{figure}[h!]
    \centering
      \subfigure[Mannix]{\includegraphics[scale=0.3]{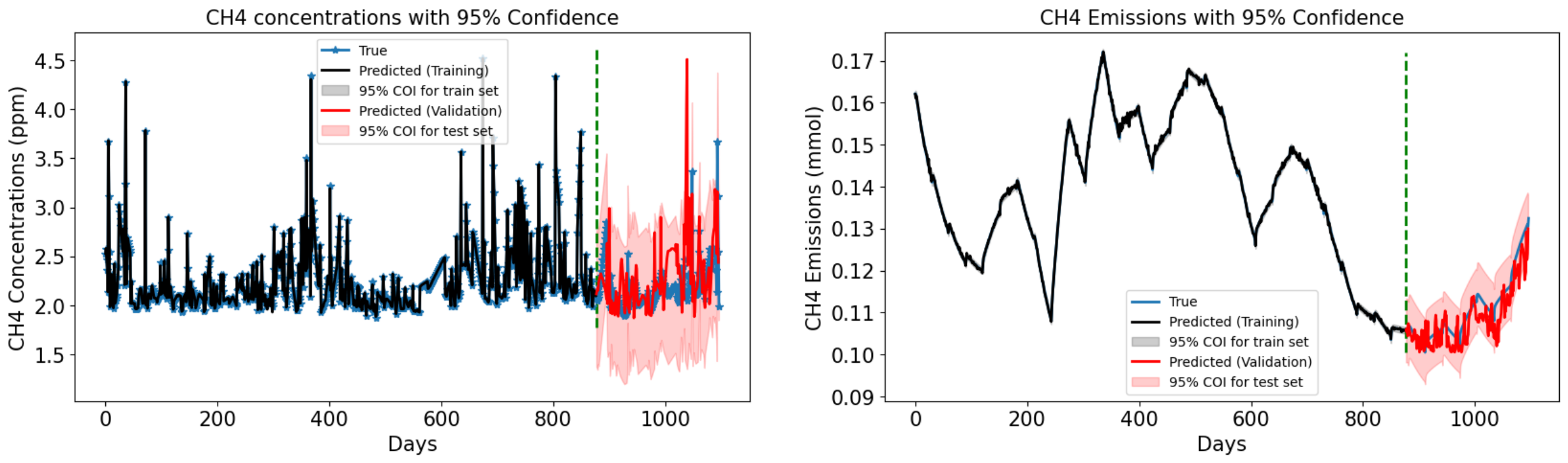}}
     \label{fig:forecast-mannix}
       \subfigure[Lower Camp]{\includegraphics[scale=0.3]{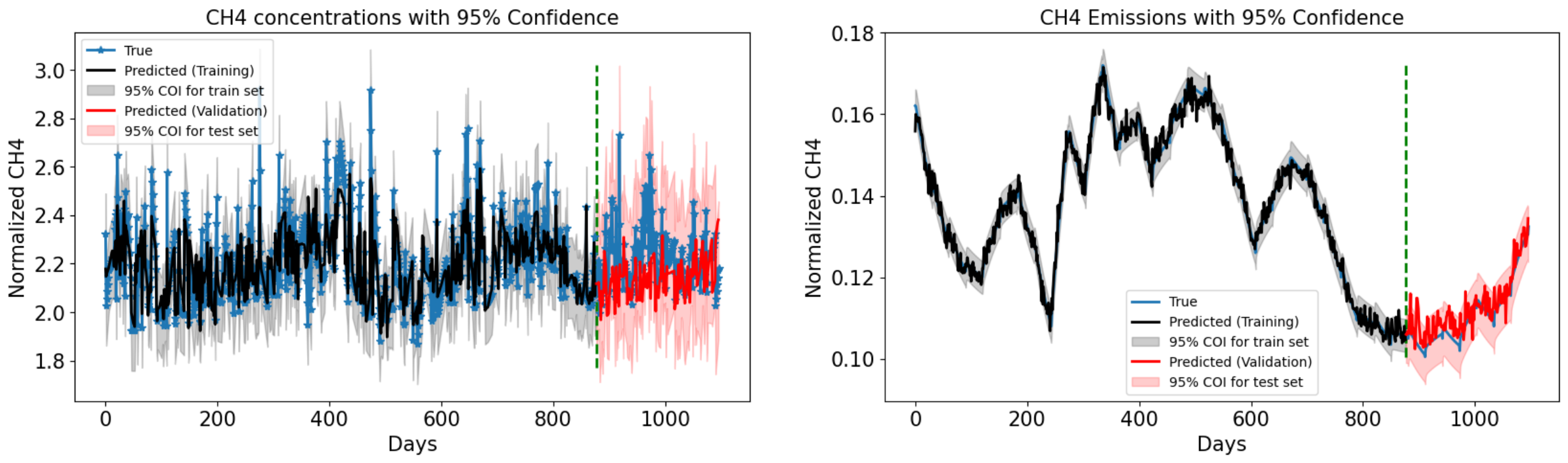}}
     \label{fig:forecast-lc}
     \subfigure[Mildred Lake]{\includegraphics[scale=0.3]{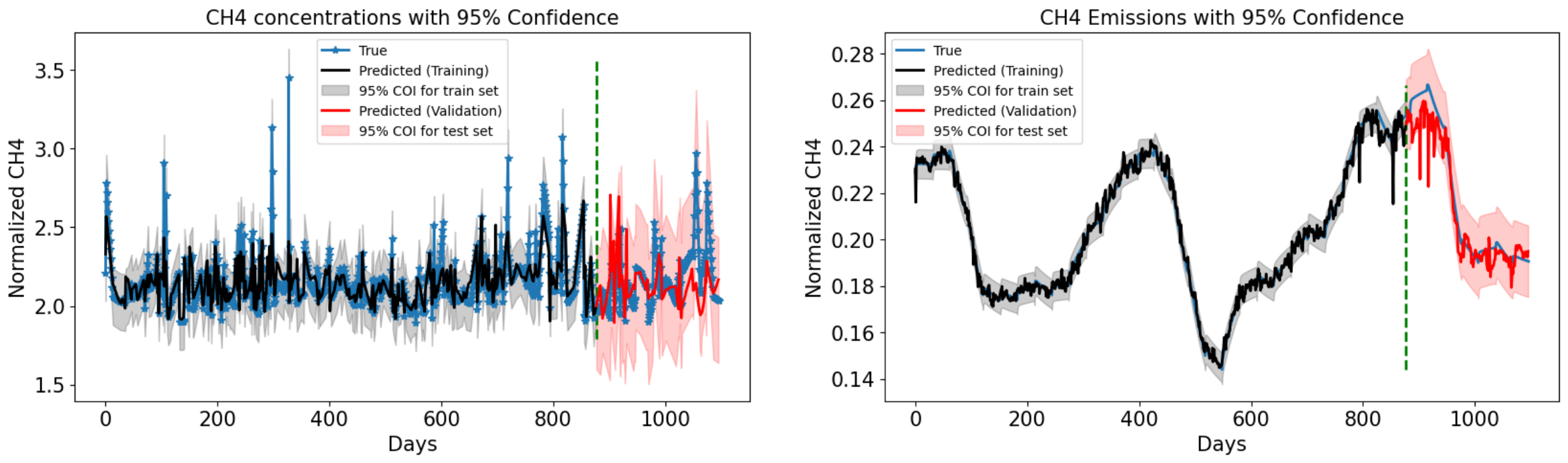}}
     \label{fig:forecast-mlsb}
    \caption{Results for predicting CH$_4$ concentrations and emissions using a neural network based representation of the dispersion/advection terms in ADM. The green dashed line indicates the data split between training and validation set. \textcolor{black}{For concentrations, the model accurately captures the seasonal nature of concentrations as well as the emissions based on given input data. The model predictions always fall within the 95\% confidence of interval.}}
    \label{fig:forcast-main}
\end{figure}

Given the mass of industrial waste in a given OSTP and the meteorological conditions around them, DIRNN is trained to predict CH$_4$ emissions from the pond and its corresponding concentration at the closest weather monitoring station. 
Figure \ref{fig:forcast-main} and Table \ref{tab:rel_errors} suggest that different representations have similar predictive power.
Other representations such as a simple Recurrent Neural Network (RNN) without dispersion constraints, Long-Short-Term-Memory (LSTM) (as defined in Section \ref{si:ml-def}) performed poorly in comparison to our proposed model. 
\begin{table}
    \centering
\begin{tabular}{|l|l|l|l|l|l|l|}
    \hline 
   \multicolumn{1}{|c|}{} &
      \multicolumn{3}{c|}{Concentration} &
      \multicolumn{3}{c|}{Emission}\\ 
      \hline
    & \small Mannix & \small LC & \small ML & \small Mannix & \small LC & \small ML \\
    \hline
    \small DIRNN$_{forward}$ & \textbf{0.5425} & \textbf{0.3101} & 0.5814 & \textbf{0.0992} & 0.1615 & \textbf{0.0517} \\
    \small iDIRNN & 0.6566 & 0.4301 & 0.5702 & 0.1085 & \textbf{0.1369} & \textbf{0.0543}\\
    \small DIRNN$_{poly}$ & \textbf{0.4624} & \textbf{0.3140} & \textbf{0.4225} & 0.5380 & 0.1977 & 0.0942\\
    \small RNN & 0.6239 & 0.4314 & 2.1859 & \textbf{0.0661} & \textbf{0.0737} & 0.7127\\
    \small LSTM & 0.6454 & 0.3914 & \textbf{0.5550} & 0.2426 & 0.2568 & 0.0711\\
    \hline
  \end{tabular}
  \caption{Average of relative training and validation errors for predicting concentrations and emissions from each station and its corresponding tailing pond. For each column (station), the two lowest errors are highlighted. \textcolor{black}{DIRNN$_{forward}$ and DIRNN$_{poly}$ denotes the training framework that uses forward dispersion model to predict concentrations from emissions. iDIRNN refers to the framework using reverse dispersion model that learns the emissions given the concentrations.}}
    \label{tab:rel_errors}
\end{table}

The results demonstrate a seasonal and diluent-dependent relationship between methane concentrations and source emissions using a data-driven ADM.
We found an increasing/decreasing trend of methane concentration and emissions based on input data consisting of atmospheric variables and hydrocarbon degradation data.
A direct connection between emissions and concentrations is established through our model indicating a rise in atmospheric CH$_4$ concentration levels right after increased emissions.
Majority of our predictions fall within the 95\% confidence bound, which was found to be narrower on training data as expected, and comparatively wider for the validation set as those predictions are based on unseen data.
Since the concentration data is noisy, we also see noise/oscillations in fitting of the emission data, which is discussed in detail in Section \ref{sec:discussion}.  

As with any machine learning algorithm, the model training is affected by the quality of training data which is what creates a difference in performance depending on the station under consideration.
For CH$_4$ concentration data station Mannix has the highest variance of 0.12 with maximum and average CH$_4$ concentrations being 4.56 ppm and 2.21 ppm respectively, followed by station Mildred Lake with a variance of 0.05, and maximum and average CH$_4$ concentrations being 3.95 ppm and 2.1 ppm respectively. Station Lower Camp has lowest variance of 0.03 with maximum and average CH$_4$ concentrations being 3.36 ppm and 2.001 ppm respectively. 
From the relative errors in Table \ref{tab:rel_errors}, it can be seen that lower the variance, better the performance of the models.
This happens as data-driven models tend to have larger biases in predicting the extreme
values, especially in data-scarce training regimes such as in this scenario.
\textcolor{black}{However, as we use other physical constraints in our framework, the model can outperform other models even with datasets with higher levels of noise and variation.}

\begin{figure}[h!]
    \centering
      \subfigure[Mannix]{
      \includegraphics[scale=0.3]{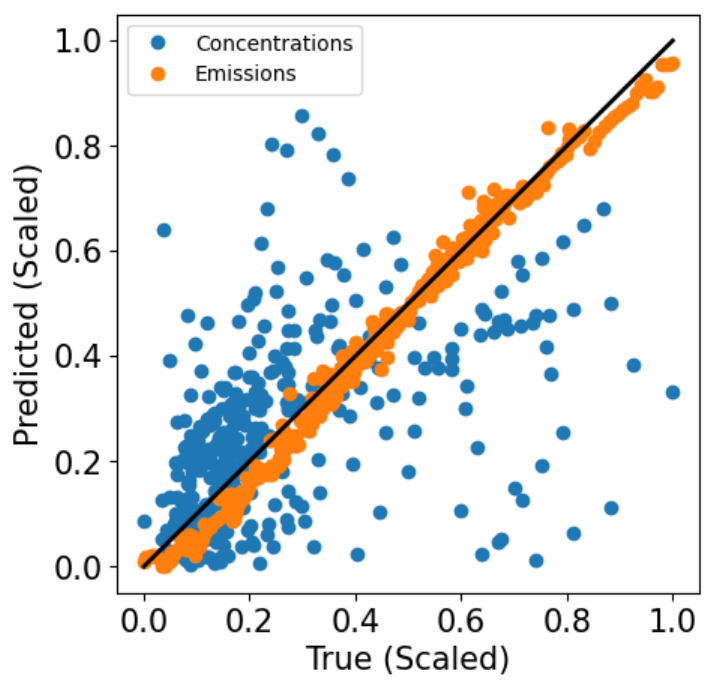}}
       \subfigure[Lower Camp]{
      \includegraphics[scale=0.3]{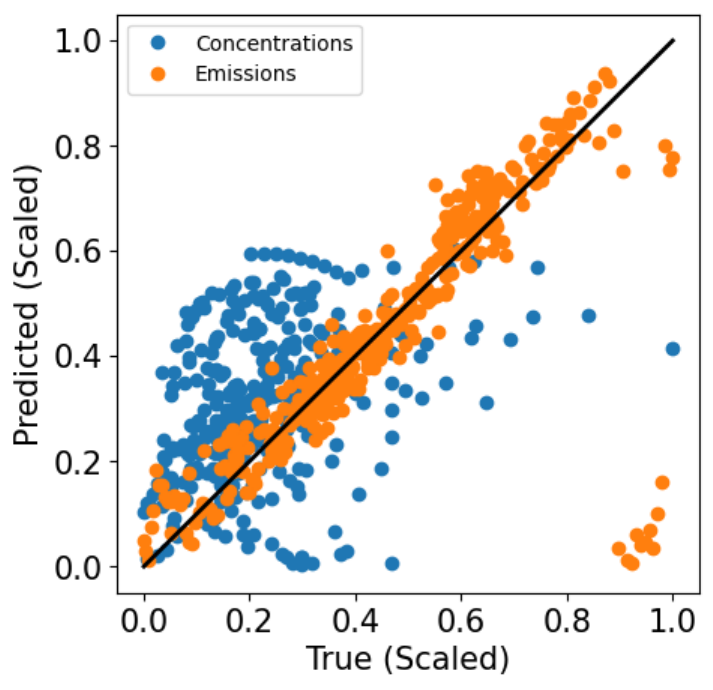}}
     \subfigure[Mildred Lake]{
     \includegraphics[scale=0.3]{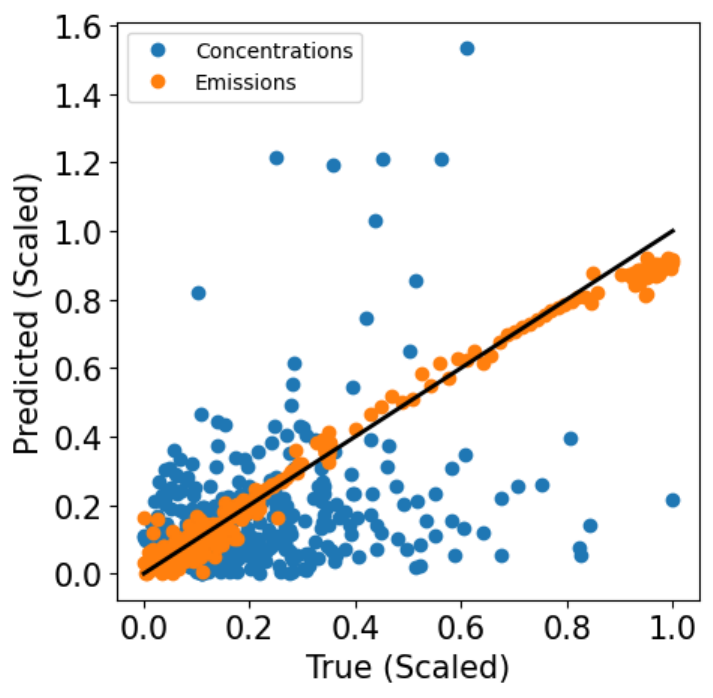}}
    \caption{Scaled values (between 0 and 1) of true versus predicted CH$_4$ concentrations and emissions in 2023 for stations Mannix, Lower Camp and Mildred Lake using a neural network representation of the dispersion/advection terms in the ADM. \textcolor{black}{The trained model accurately forecasts concentrations and emissions for one year when trained on historical data, suggesting future trends based on different levels of input data. }}
    \label{fig:2023-predict}
\end{figure}

To test our model's long term CH$_4$ concentration predictive capacity, we obtain predictions for the year 2023 and compare it with the observed data. 
Figure \ref{fig:2023-predict} shows that the trained model accurately estimates emissions from OSTPs upto one year ahead.
For predicting concentrations, the model can suggest future trends, with better data fitting when the true data is closer to the mean CH$_4$ levels.
However, the variation is larger (for higher recorded observation) since the real-time data is noisy and the model ignores extreme large values as outliers. 
Our results indicate that with current levels of oil sands activity and similar meteorological parameter readings (temperature, humidity, etc), there will be no improvement in CH$_4$ concentrations in the region.

\subsection{Identification of CH$_4$ Sources around Weather Stations}
Due to the Athabasca region being located so further up in the North, a lot of the existing remote sensing methods for continuous monitoring of emissions are not directly applicable. 
The satellite data is often of poor quality (due to low sunlight, especially during winters and/or cloud cover). 
Other data measurement techniques (example, airplanes or drones) generally do not differentiate between multiple sources of emissions and are mostly expensive to carry out on a regular basis). 
Thus, to build a CH$_4$ emission monitoring method, we train the proposed model using the reverse formulation of ADM to estimate daily emissions $360$ degrees around a selected weather station.
Once trained, we replace the input $u_{\bar{\Theta}}$ in Eq. \eqref{eq:DIRNN_rev} with true CH$_4$ concentration data from weather monitoring stations to obtain daily emission estimates from the tailing ponds. 
\begin{figure*}[!h]
\centering
\includegraphics[scale = 0.4]{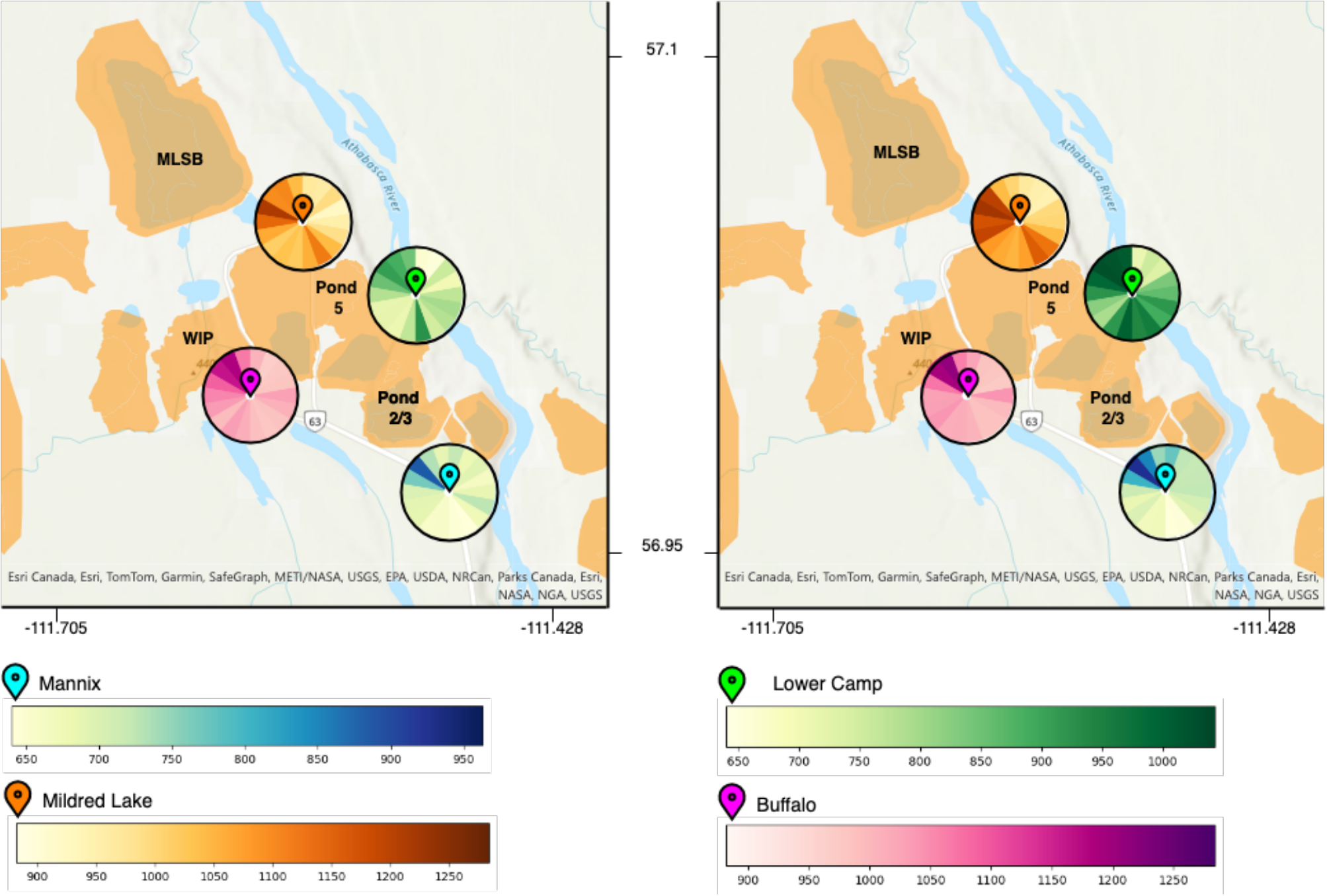}
\caption{Estimation of yearly CH$_4$ emissions (in t) from all wind directions around each station for 2020 (left) and 2023 (right). 
\textcolor{black}{Estimated emissions are highest from the direction of OSTPs, with MLSB emitting 4500t per year and Pond 2/3 emitting more than 850 tonnes (t) per year every 20 degree interval. Abandoned lakes such as WIP and Pond 5 also emits more than 2000t of CH$_4$ per year.}
Please note that the round plots are for demonstration purpose and are not to scale. Latitude and longitude coordinate ranges given in $X$ and $Y$ axis.}\label{fig:track-ch4}
\end{figure*}
Daily predicted emissions are summed up to obtain cumulative emissions for each year between 2020 to 2023 and is given as a radial plot in Figure \ref{fig:emm_track}.  
To compare emission levels over a period of three years, we plot the emissions for the year 2020 and 2023 in Figure \ref{fig:track-ch4}.
The two major reported OSTPs (MLSB and Pond 2/3) and abandoned pond/EPLs (Pond 5 and WIP) are marked on the map. 
For weather monitoring stations in the vicinity of inactive ponds, we used a trained model from another appropriate station to get emission estimates.
For example, since WIP lake (an inactive pond) near station Buffalo is owned by Syncrude, we can use a model trained with Syncrude's data to get emission estimates around it. Here we used the model trained for the Mildred Lake station.
For each station we see the estimated emissions are highest from the direction of tailing ponds. Both Mannix and Lower Camp indicate that CH$_4$ emissions from Pond 2/3 are more than 850 tonnes (t) per year every 20 degree interval. 
Each year, station Mildred Lake captures more than 4500 t of CH$_4$ emissions from the direction of MLSB.
The other inactive ponds (WIP and Pond 5) are often ignored as a significant sources of CH$_4$ emissions as there is no inflow of diluents. 
However, using the dataset of station Buffalo, iDIRNN estimates more than 2000 t of emissions coming from the direction of WIP lake and about 1900 t of CH$_4$ from Pond 5.
These significant levels of emissions from Pond 5 can also be cross-verified with the results obtained for stations Lower Camp and Mildred Lake. 
While perfect sectoral disaggregation of sources is challenging, the choice of weather stations (close to OSTPs with no other CH$_4$ source in between), use of radial wind directions as established in prior works and the fact that CH$_4$ is generally measured within 5-15 $m$ from the ground (indicating that CH$_4$ does not disperse very far from its source, thus the concentration recorded at the station has a source nearby) gives a strong indication of OSTPs being CH$_4$ emission sources.

Upon comparing the plots for the years 2020 (left) and 2023 (right) in Figure \ref{fig:track-ch4}, we see that there has been a significant increase in emissions over the three years by at least 100 t in each of the 20 degree wind direction interval near the tailing ponds. Both MLSB and Pond 2/3 have started contributing higher amounts of CH$_4$. 
For year 2023, we also find elevated emissions for both Mannix and Lower Camp with wind directions 20-140 degrees and 60-140 degrees respectively, coming from OSTPs (such as Pond 8 owned by Suncor or similar sources in the area) south-east of the stations across the Athabasca river.
Some other directions with lower emission levels of about 700-900 t (West of Mannix, South-East of Buffalo, SW of Lower Camp), all correspond to either in-situ facilities, industrial activity or inactive tailing ponds. 
Note that, the confidence interval for the emission predictions were similar to that obtained earlier in Figure \ref{fig:forcast-main}. However, since visualization of the interval is challenging in a radial plot, we have not included them in Figure \ref{fig:track-ch4}. 

\section{Discussion}\label{sec:discussion}
 Atmospheric CH$_4$ can be affected by natural as well as anthropogenic factors. 
 In order to build a reliable framework, DIRNN was trained by connecting all possible dynamics connected to CH$_4$ emissions from OSTPs and the air quality in the region: MM of hydrocarbon degradation \citep{pablo,kong2019second}, ADMs for CH$_4$ dispersion \citep{stockie2011mathematics} and atmospheric data that drives atmospheric diffusion and advection.
The two OSTPs used for dataset building are appropriate choices for representing majority of the OSTPs in the region mainly because of two reasons: (i) both of these ponds are not only big in terms of area but have also been previously found to emit much larger amounts of CH4 in comparison to other ponds; (ii) an important aspect of CH$_4$ emissions from OSTPs is based on diluent composition which is determined by the owner company. Both these ponds take into account two of the major companies: Syncrude (now Suncor) and Suncor for our studies. 
These two ponds have been used in prior works frequently to study the extent of OSTP emissions \citep{you2020methane,small2015emissions}.

The model performances are compared based on both, plots and relative errors. 
Table \ref{tab:rel_errors} shows that for predicting concentrations, different parameterization of the $\nabla.J$ term can have similar predictive capabilities \citep{andoni2014learning}. 
The differences in the model performances between stations are attributed to the quality and quantity of data available for building the dataset.
External factors such as the proximity of OSTPs to the weather stations, height of measurement sensors, etc. can affect data quality.
For example, hypothetically if one of the stations were to be on a cliff (e.g. WBEA Stony Mountain, not used in the paper) it is more likely to be affected by high wind speeds \citep{solano2021correlation}, thus leading to a noisy dataset. 
Since the concentration data is noisy, we also see noise/oscillations in fitting of the emission data \textcolor{black}{(also knonw as Gibbs phenomenon oscillations in literature) which is common in literature with model learning using noisy data \citep{berry2016semiparametric}. 
Underlying model can be successfully recovered even with noisy data as long as the variation of noise is low \citep{tran2017exact} and the fitted function lies within the confidence of interval, thus controlling range within which the true values are expected to lie. Previous works have showed that Gibbs phenomenon does not majorly affect global generalizability of model approximation, and that different techniques of filtering illogical approximations (for example, negative values in a positive function approximation) is acceptable \citep{gottlieb1997gibbs, berry2016semiparametric}.} 
The CH$_4$ concentration data collected for station Mannix has the highest variance of 0.12, followed by station Mildred Lake with a variance of 0.05, and then station Lower Camp with a variance at 0.03. 
From the table of relative errors (Table \ref{tab:rel_errors}), it can be seen that a lower the variance improves the performance of the models.
This happens as data-driven models tend to have larger biases in predicting the extreme
values, especially in data-scarce training regimes such as in this scenario.
Superior model performance for the station Lower Camp can also be attributed to its restricted range of wind direction during the dataset building stage, making the input dataset more less noisy in comparison to other stations.  
Note that since the ponds are owned by different companies, the hydrocarbon degradation dynamics can differ based on the chemical composition of diluents used, as well as the accuracy of the data reported by the companies. 
Although emission predictions are slightly noisy, it does not affect the generalization power of our trained model as it preserves the overall trends on unseen data for both the quantities.

In the reverse formulation of the model, we use the trained model to monitor emissions from all the direction around each weather monitoring station. 
We consider the real concentration data along with the weather parameters inside the trained model to track emissions and possible sources (replace the input $u_{\bar{\Theta}}$ in Eq. \eqref{eq:DIRNN_rev} with true CH$_4$ concentration data). 
Thus, for sources such as OSTPs (or other sources emitting CH$_4$ from diluent degradation) the model can be used as a tool to monitor emissions around weather stations.
Since abandoned ponds and EPLs are often ignored as sources, our goal is also to monitor these tailings for emissions. 
A thorough analysis of emissions over four years in Figure \ref{fig:emm_track} shows that emissions from active OSTPs are not only increasing every year, some of the other inactive ponds/EPLs such as WIP and Pond 5 could also be significant sources of CH$_4$ emissions.
\textcolor{black}{Overall, our estimates show an underestimation in official reports \citep{nir,emmreports} atleast by a factor of three.}

\subsection{Comparison to Other Works and Data}
In order to put our emission estimates into the context of existing literature, we take a look at the trend of emission estimates from previous works in Table \ref{tab:previous_est}. 
Note that since these studies were conducted across a span of last ten years, the emission estimates vary. 
In fact, they may also vary for different methods within the same year depending on the estimation method used.
\begin{table}[h!]
    \centering
    \begin{tabular}{|c|c|c|}
    \hline
        Name/Method & Year & Emission (t/y) \\
        \hline
  \citep{small2015emissions} & 2015 & 2657.2  \\
         \citep{you2020methane} & 2016 & 3876.6 \\
         \citep{you2020methane} & 2018 & 11344.2\\
          Windtrax \citep{crenna2016introduction} & 2021 & 5518 \\
          DIRNN & 2020-2023 & 3498.75\\
         \hline
    \end{tabular}
    \caption{Comparison of CH$_4$ emission estimates for Pond 2/3 obtained from prior works and inverse dispersion modeling software WindTrax. Note that all the estimates are made for different years and may not be comparable directly. Units from original works are converted to tonnes per year for the 2.8 $km^2$ OSTP Pond 2/3.}
    \label{tab:previous_est}
\end{table}
We include estimates using WindTrax \citep{crenna2016introduction}, a Lagrangian stochastic particle based model and calculated with parameters in \citep{you2020methane}. Note that using WindTrax with limited weather station input data for 2020 - 2023 (such as from Lower Camp) leads to highly unstable results, with unrealistic outputs varying over $10^5$t per year. Thus, the outputs from those years are excluded from Table \ref{tab:previous_est}.  
Our estimates are based on emissions averaged over the four years 2020-2023 calculated from the station Lower Camp (or Mannix; both give similar estimates). CH$_4$ emissions using iDIRNN fall within the overall range of what is suggested by previous works. Our estimations of 3498.75 t/y are close to results showed by most of the estimations in \citep{you2020methane,small2015emissions}.
Note that due to different time periods, it may be hard to pinpoint which technique is the most accurate one. 

\textcolor{black}{The results underscore the need to increase efforts in accurately estimating CH$_4$ emissions and concentrations owing to its environmental and health impacts.  
Oil sands activities contribute significantly to GHG emissions and in particular are regarded as sources of pollution \citep{schindler2014unravelling,liggio2016oil,yu2023tailings}. 
From the perspective of environmental effects, mining and extraction of oil sands is directly associated with deforestation and release of sulfur oxides, nitrogen oxides, hydrocarbons, and fine particulate matter, etc.
Further, CH$_4$ reacts with hydroxyl radical leading to the formation of ground-level ozone, which is a harmful and toxic air pollutant.
While current levels of CH$_4$ concentration do not have direct health impact ($\sim$ 2 ppm - 4.5 ppm), consequent displacement of oxygen, and ozone formation can cause symptoms such rapid heart rate, fatigue and other health affects from lack of oxygen and increased air pollution \citep{adgate2014potential,healthch4impact}. }

\textcolor{black}{Our study also aims to extend the growing body of research emphasizing on the role of machine learning in promoting sustainable practices, and efficient resource utilization.
By integrating physics-informed deep learning into the monitoring and estimation of methane emissions from oil sands tailings ponds, our work contributes to the advancement of cost-effective environmental monitoring systems, offering tools for improving accountability, resource management, and regulatory compliance within the oil and gas sector industries.
By making emission estimation scalable, accurate, and physically grounded, our methodology enhances decision-making capacity for governments, industries, and communities alike, promoting a fairer distribution of environmental responsibilities and better-informed climate governance. }

\subsection{Model Uncertainties, Limitations and Future Work} \textcolor{black}{
Being significantly data dependent, the training and results of our framework are limited to quality and quantity of available data. The data can be highly variable and prone to noise. For example, calibration errors, human recording errors, etc. The averaging step of processing the atmospheric data helps to ensure that noise in the data is reduced. 
The hybrid machine learning model also has a possibility of overfitting and unstable training. To avoid this, the results were based on the median values of multiple rounds of model training and validation with carefully tuned hyperparameters.
Also note that the same trained model may not be applicable to other OSTPs without retraining. A model trained on one OSTP (and weather station) can be used on another location only if all the following conditions are met: (a) the OSTP under consideration is owned by the same company ensuring that the chemical composition of diluents remain the same; (b) it is located in a region with similar distribution of input parameters (similar weather conditions and/or diluent composition); (c) close proximity to the original OSTP and the weather monitoring station.
For example, in order to get results for OSTPs owned by Shell or CNRL, we would need to retrain the model as (a) diluent composition of these companies are different; (b) they are far from the OSTPs we trained our model on. 
Since the model training is quick, if data is available, we would always recommend retraining to get the most accurate and reliable results. 
Future potential of this work is vast and includes improving MMs through advanced lab experiments, incorporating data from all directions around the weather stations to account for all possible CH$_4$ emitting sources and modeling the affects of these emissions on climate change. }

\section{Conclusions}\label{sec:conclusion}
In this paper, we developed a hybrid machine learning approach for predicting methane emissions and concentrations jointly. Our model formulation was based on learning the atmospheric methane concentrations using
data obtained from weather monitoring stations subject to atmospheric
dispersion models. We use a recurrent neural network style formulation to learn both, methane concentrations and the unknown
functions in the constraints. Our input dataset included measurements from weather monitoring stations
located within 4 km of active OSTPs and emission data obtained from solving different methanogenesis
models.
\textcolor{black}{Along with providing accurate forecasts for CH$_4$ emissions and concentrations, our model results indicated significant emissions from both active and abandoned OSTPs, suggesting atleast three times underestimation in existing official reports.
Our results using the proposed approach were consistent with existing estimates and outperformed other classical machine learning approaches.
As a part of future work,
we plan to incorporate sources from all the wind directions into the framework by either using additional
constraints based on existing mathematical models of various methane sources or by incorporating aspects
of remote sensing to gather data of methane emissions around the OSTPs.}

\section*{Acknowledgments}
The authors would like to acknowledge the and thank the Natural Sciences and Engineering Research Council of Canada (NSERC) for funding through an NSERC Alliance Missions grant on anthropogenic greenhouse gas research.

\bibliographystyle{elsarticle-harv}

\appendix
\section{Definitions}\label{si:ml-def}

\begin{enumerate}
    \item A \textit{neural network} is a parameterized function composed of a sequence of affine transformations followed by nonlinear activations. Formally, a feedforward neural network with $L$ layers can be expressed as:

\begin{equation}
\begin{aligned}
    \mathbf{h}^{(0)} &= \mathbf{x} \\
    \mathbf{h}^{(l)} &= \sigma\left( \mathbf{W}^{(l)} \mathbf{h}^{(l-1)} + \mathbf{b}^{(l)} \right), \quad \text{for } l = 1, \dots, L-1 \\
    \mathbf{y} &= \mathbf{W}^{(L)} \mathbf{h}^{(L-1)} + \mathbf{b}^{(L)}
\end{aligned}
\end{equation}

where $\mathbf{x} \in \mathbb{R}^{n}$ is the input vector, $\mathbf{h}^{(l)}$ is the hidden representation at layer $l$, $\mathbf{W}^{(l)}$ and $\mathbf{b}^{(l)}$ are the weight matrix and bias vector at layer $l$, $\sigma(\cdot)$ is a nonlinear activation function (e.g., ReLU, sigmoid), $\mathbf{y}$ is the output of the network.

\item A \textit{Recurrent Neural Network} (RNN) is a type of neural network designed for modeling sequential data by maintaining a hidden state that captures information from previous time steps. The computations at each time step $t$ are given by:

\begin{equation}
\begin{aligned}
    \mathbf{h}_t &= \phi\left( \mathbf{W}_{xh} \mathbf{x}_t + \mathbf{W}_{hh} \mathbf{h}_{t-1} + \mathbf{b}_h \right) \\
    \mathbf{y}_t &= \mathbf{W}_{hy} \mathbf{h}_t + \mathbf{b}_y
\end{aligned}
\end{equation}

where $\mathbf{x}_t \in \mathbb{R}^{n}$ is the input vector at time step $t$, $\mathbf{h}_t \in \mathbb{R}^{m}$ is the hidden state at time $t$, $\mathbf{y}_t \in \mathbb{R}^{k}$ is the output at time $t$, $\mathbf{W}_{xh}$ is the input-to-hidden weight matrix, $\mathbf{W}_{hh}$ is the hidden-to-hidden (recurrent) weight matrix, $\mathbf{W}_{hy}$ is the hidden-to-output weight matrix, $\mathbf{b}_h$, $\mathbf{b}_y$ are bias vectors, $\phi(\cdot)$ is an activation function, typically $\tanh$ or ReLU.

The hidden state $\mathbf{h}_t$ allows the RNN to retain memory of previous inputs, enabling the modeling of temporal dependencies in sequences.

\item A \textit{Long Short-Term Memory} (LSTM) network is a type of recurrent neural network (RNN) designed to capture long-range dependencies by introducing memory cells and gating mechanisms. The operations for a single LSTM cell at time step $t$ are defined as:

\begin{equation}
\begin{aligned}
    \mathbf{i}_t &= \sigma\left( \mathbf{W}_{xi} \mathbf{x}_t + \mathbf{W}_{hi} \mathbf{h}_{t-1} + \mathbf{b}_i \right) \quad &\text{(input gate)} \\
    \mathbf{f}_t &= \sigma\left( \mathbf{W}_{xf} \mathbf{x}_t + \mathbf{W}_{hf} \mathbf{h}_{t-1} + \mathbf{b}_f \right) \quad &\text{(forget gate)} \\
    \mathbf{o}_t &= \sigma\left( \mathbf{W}_{xo} \mathbf{x}_t + \mathbf{W}_{ho} \mathbf{h}_{t-1} + \mathbf{b}_o \right) \quad &\text{(output gate)} \\
    \mathbf{g}_t &= \tanh\left( \mathbf{W}_{xg} \mathbf{x}_t + \mathbf{W}_{hg} \mathbf{h}_{t-1} + \mathbf{b}_g \right) \quad &\text{(cell candidate)} \\
    \mathbf{c}_t &= \mathbf{f}_t \odot \mathbf{c}_{t-1} + \mathbf{i}_t \odot \mathbf{g}_t \quad &\text{(cell state update)} \\
    \mathbf{h}_t &= \mathbf{o}_t \odot \tanh(\mathbf{c}_t) \quad &\text{(hidden state output)}
\end{aligned}
\end{equation}

where $\mathbf{x}_t$ is the input vector at time $t$, $\mathbf{h}_{t}$ is the hidden state (output) at time $t$, $\mathbf{c}_t$ is the cell state (memory) at time $t$, $\mathbf{i}_t$, $\mathbf{f}_t$, $\mathbf{o}_t$ are the input, forget, and output gates, respectively, $\mathbf{g}_t$ is the candidate cell state, $\sigma(\cdot)$ is the sigmoid activation function, $\odot$ denotes element-wise (Hadamard) product, $\mathbf{W}_{x\ast}$ and $\mathbf{W}_{h\ast}$ are weight matrices, and $\mathbf{b}_\ast$ are bias vectors.

\item A \textit{Physics-Informed Neural Network} (PINN) \citep{raissi2019physics} is a neural network model trained not only to fit observed data, but also to satisfy the underlying physical laws (usually expressed as partial differential equations, PDEs). 
Let $\mathcal{N}_\theta(\mathbf{x})$ be a neural network parameterized by weights $\theta$ that approximates the solution $u(\mathbf{x})$ of a PDE defined over a domain $\Omega$. The governing PDE can be written as:

\begin{equation}
    \mathcal{F}(u(\mathbf{x}), \nabla u(\mathbf{x}), \nabla^2 u(\mathbf{x}), \dots) = 0, \quad \mathbf{x} \in \Omega,
\end{equation}

where $\mathcal{F}$ is a differential operator representing the physical law.

The training objective of a PINN involves minimizing a composite loss function:

\begin{equation}
    \mathcal{L}(\theta) = \mathcal{L}_{\text{data}} + \lambda \, \mathcal{L}_{\text{physics}},
\end{equation}

where $\mathcal{L}_{\text{data}} = \frac{1}{N_d} \sum_{i=1}^{N_d} \left\| \mathcal{N}_\theta(\mathbf{x}_i) - u_i^{\text{obs}} \right\|^2$ is the supervised data loss at observed data points $\{(\mathbf{x}_i, u_i^{\text{obs}})\}$, $\mathcal{L}_{\text{physics}} = \frac{1}{N_f} \sum_{j=1}^{N_f} \left\| \mathcal{F}\left(\mathcal{N}_\theta(\mathbf{x}_j)\right) \right\|^2$ is the physics-based loss over collocation points $\{\mathbf{x}_j\}$ in the domain, $\lambda$ is a weighting coefficient balancing the two loss components.

PINNs leverage automatic differentiation to compute derivatives of $\mathcal{N}_\theta(\mathbf{x})$ required in $\mathcal{F}$, allowing the model to learn solutions that respect both empirical data and the governing physics.

\end{enumerate}

\section{Model Framework}\label{sec:data_si}

The framework is based on a simple relationship between emissions and concentrations: hydrocarbons in tailings ponds directly affect emissions which affects concentrations. Since the process of methanogenesis takes place under the OSTPs, majority of the atmospheric variables have no effect on hydrocarbon degradation. 
However, once released from the source, emissions along with other atmospheric parameters directly affect concentration levels. 
Thus, the input to the proposed framework includes various parameters that directly or indirectly affect atmospheric methane concentrations. 
The model considers three types of input data: (i) $\mathbf{x}_{dil}$ denoting the degradation of hydrocarbons in OSTPs and obtained from solving MMs in literature; (ii) $\mathbf{x}_{atm}$ representing atmospheric parameters such as ambient temperature, wind speed, wind direction, solar activity, etc; and (iii) time vector $\mathbf{t}$. 
These three inputs $\mathbf{x}_{dil}$, $\mathbf{x}_{atm}$, and $\mathbf{t}$ together form the input $\mathbf{x}$ and are used to define the minimization problem.
While the proposed method can be combined with a variety of machine learning architectures and constraints, we propose three general formulations of DIRNN, each with its own advantages and applications. Table \ref{tab:compare} gives all the variations as well as some other models that were used for comparison. Their architectures and outputs are described along with a comment on if the minimization problem arising from them are constrained or not. 
\begin{table}
    \centering
    \begin{tabular}{|c|c|c|}
    \hline
        Model name & Formula &  Constrained \\
        \hline
         DIRNN$_{\text{Forward}}$ &  \makecell{$ \text{CH}_4^{(1)} = \bar{\Theta}_4 \sigma\left(\bar{\Theta}_3\sigma\left(\bar{\Theta}_2\sigma\left(\bar{\Theta}_1\mathbf{x}\right)\right)\right)$\\
    $\text{CH}_4^{(2)} = \text{Grad}_t\left(\text{CH}_4^{(1)}\right) + \hat{\Theta}_3\sigma\left(\hat{\Theta}_2\sigma\left(\hat{\Theta}_1\left[
        \text{CH}_4^{(1)} ,
        \mathbf{x}_{atm}]
    \right]\right)\right)$} & {\color{blue}\checkmark}\\
    \hline
         iDIRNN & \makecell{$\text{CH}_4^{(1)} =\bar{\Theta}_3\sigma\left(\bar{\Theta}_2\sigma\left(\bar{\Theta}_1\mathbf{x}\right)\right)$\\
    $\text{CH}_4^{(2)} = \hat{\Theta}_3\sigma\left(\hat{\Theta}_2\sigma\left(\hat{\Theta}_1\left[
        \text{CH}_4^{(1)} ,
        \mathbf{x}_{atm}
    \right]\right)\right)$}   & {\color{blue}\checkmark} \\
    \hline
         DIRNN$_{\text{Poly}}$ & \makecell{$\text{CH}_4^{(1)} = \bar{\Theta}_4 \sigma\left(\bar{\Theta}_3\sigma\left(\bar{\Theta}_2\sigma\left(\bar{\Theta}_1\mathbf{x}\right)\right)\right)$\\
    $\text{CH}_4^{(2)} = \text{Grad}_t\left(\text{CH}_4^{(1)}\right) + \hat{\Theta}_1 \mathcal{P}\left(\left[
        \text{CH}_4^{(1)} ,
        \mathbf{x}_{atm}]
    \right]\right)$}  & {\color{blue}\checkmark} \\
    \hline
      RNN & 
    \makecell{$\text{CH}_4^{(1)} =\bar{\Theta}_4 \sigma\left(\bar{\Theta}_3\sigma\left(\bar{\Theta}_2\sigma\left(\bar{\Theta}_1\mathbf{x}\right)\right)\right) $\\
     $\text{CH}_4^{(2)} = \hat{\Theta}_3\sigma\left(\hat{\Theta}_2\sigma\left(\hat{\Theta}_1
        \text{CH}_4^{(1)}\right)\right) $} & {\color{blue}\checkmark}\\
        \hline
         NN & $\begin{bmatrix}
             \text{CH}_4^{(1)}\\
              \text{CH}_4^{(2)}
         \end{bmatrix}$ = $\Theta_5\sigma\left(\Theta_4\sigma\left(\Theta_3\sigma\left(\Theta_2\sigma\left(\Theta_1\
        \mathbf{x} \right)\right)\right)\right)$ &  {\color{black}$\boldsymbol{\times}$} \\
        \hline
         LSTM & $\begin{bmatrix}
             \text{CH}_4^{(1)}\\
              \text{CH}_4^{(2)}
         \end{bmatrix} = \text{LSTM}(\mathbf{x})$ &  {\color{black}$\boldsymbol{\times}$}\\
         \hline
    \end{tabular}
     \caption{Different models used for comparing predictions of methane emissions and concentrations.}
    \label{tab:compare}
\end{table}

\section{Additional Results}
In this section we discuss the performance of our proposed model and its variants. 
We test our model's performance on each of the three datasets and compare it with some alternative models (see Table \ref{tab:compare}). 
Three more architectures are also trained for benchmarking. 
We use another constrained version of the optimization problem where we consider a non-linear relationship between only CH$_4^{(1)}$ and CH$_4^{(2)}$. Note that we do not add atmospheric variables while computing $F$. Given the nature of the architecture, it is very similar to a RNN, although we do not use previous timesteps and thereby ignoring the `recurrent' part. We refer to it as RNN indicating it to be a similar to a simplified RNN version. The second model is a fully connected deep neural network (NN) that directly computes CH$_4^{(1)}$ and CH$_4^{(2)}$ as the outputs. Finally, we also include results from using a Long-Short-Term-Memory (LSTM) architecture. 

The model performances were assessed using two criteria: (i) average of the relative $\ell_2$ error on training and validation set; (ii) plots of true and predicted outputs. The relative error is computed for both concentrations and emissions. The formula for computing the relative error ($RE$) is given by
\begin{equation}
    RE (y_{true},y_{pblack}) = \dfrac{\|y_{true} - y_{pblack}\|_2}{\|y_{true}\|_2}
\end{equation}
where $y_{true}$ and $y_{pblack}$ denote the true and predicted outputs, respectively. The two-fold assessment of using both relative errors and plots helps us to compare how well each method can generalize the data on unseen data, especially in cases when their errors are comparable. The table of relative errors is given below in Table \ref{tab:rel_errors}.

\subsection{Forecasting Methane Emissions and Concentrations}
\subsubsection{Mannix}\label{sec:mann}
\begin{figure}[h]
    \centering
    \begin{subfigure}
     \centering
      \includegraphics[scale=0.2]{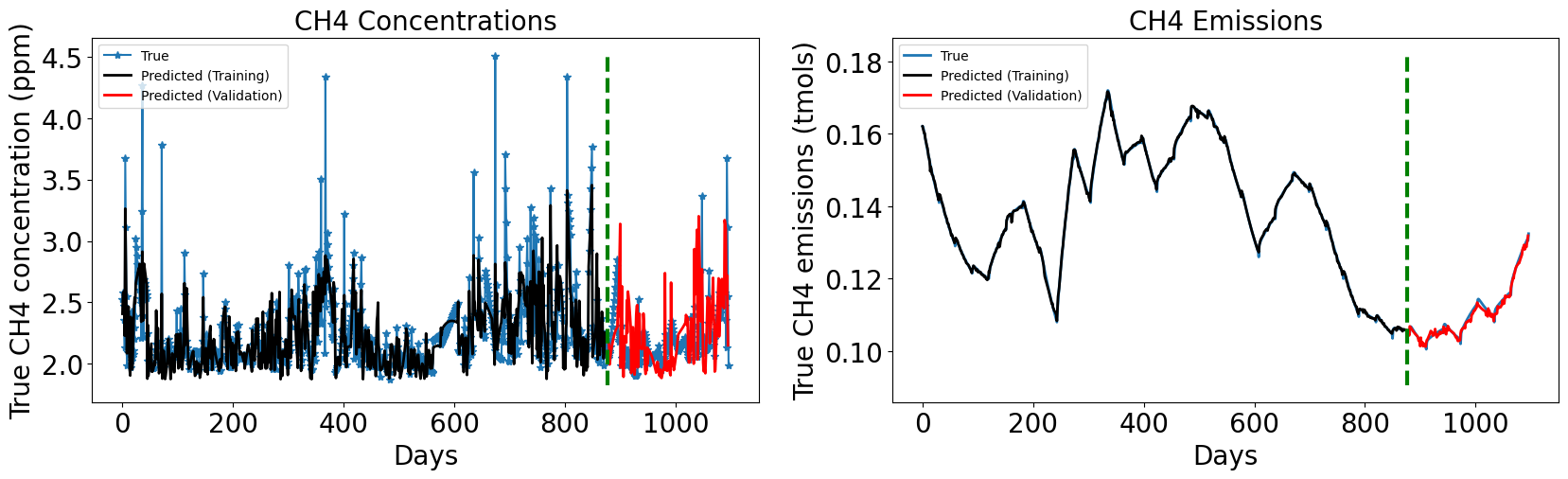}
      \caption{Forward ADM Constraint}
     \label{fig:Mannix_for}
 \end{subfigure}\,\,\,\,\,\,\,\,\,\,
 \begin{subfigure}
     \centering
      \includegraphics[scale=0.2]{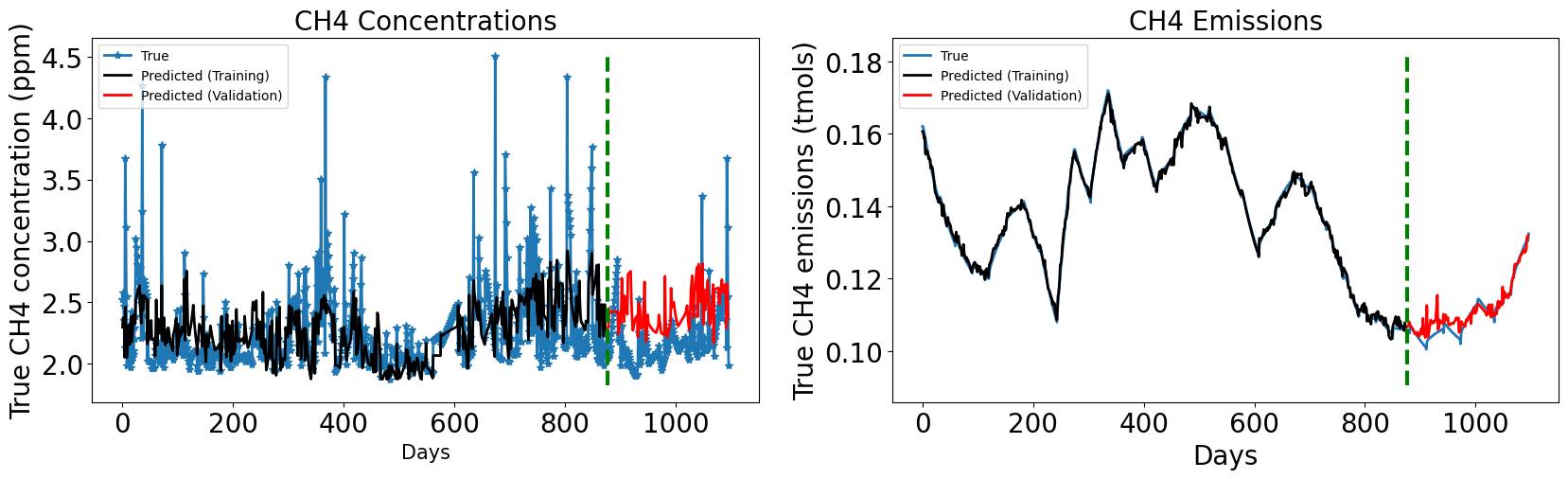}
      \caption{Reverse ADM Constraint}
     \label{fig:Mannix_rev}
 \end{subfigure}
 \begin{subfigure}
     \centering
      \includegraphics[scale=0.2]{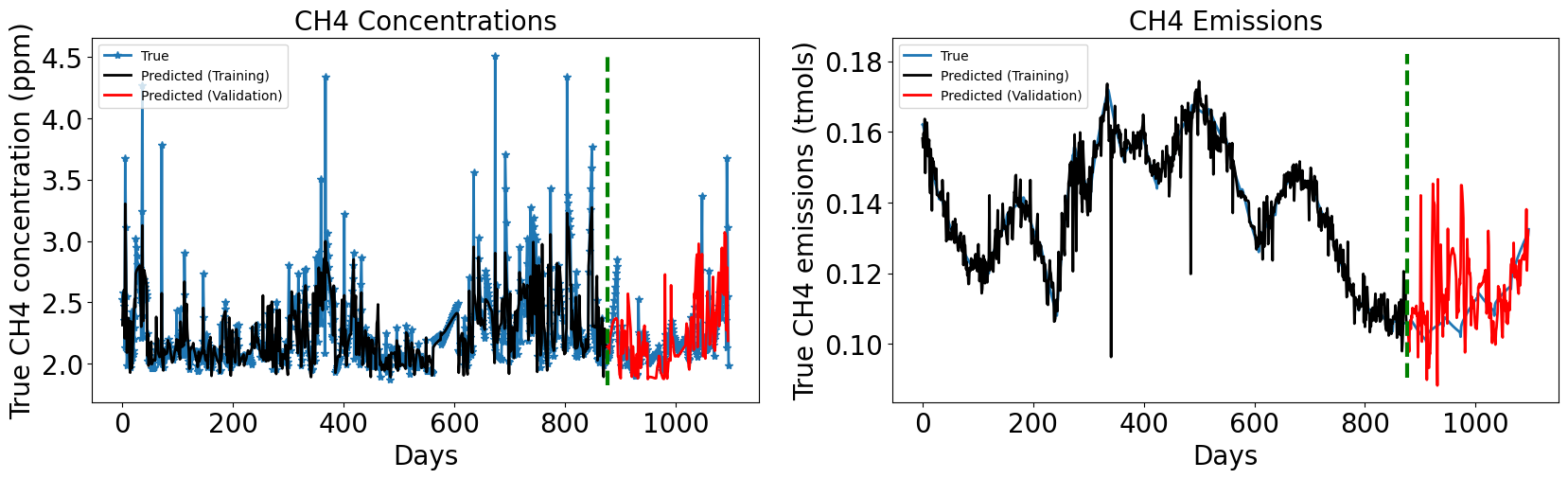}
      \caption{Polynomial Representation of ADM Constraint}
     \label{fig:Mannix_spapoly}
 \end{subfigure}
 \begin{subfigure}
     \centering
      \includegraphics[scale=0.2]{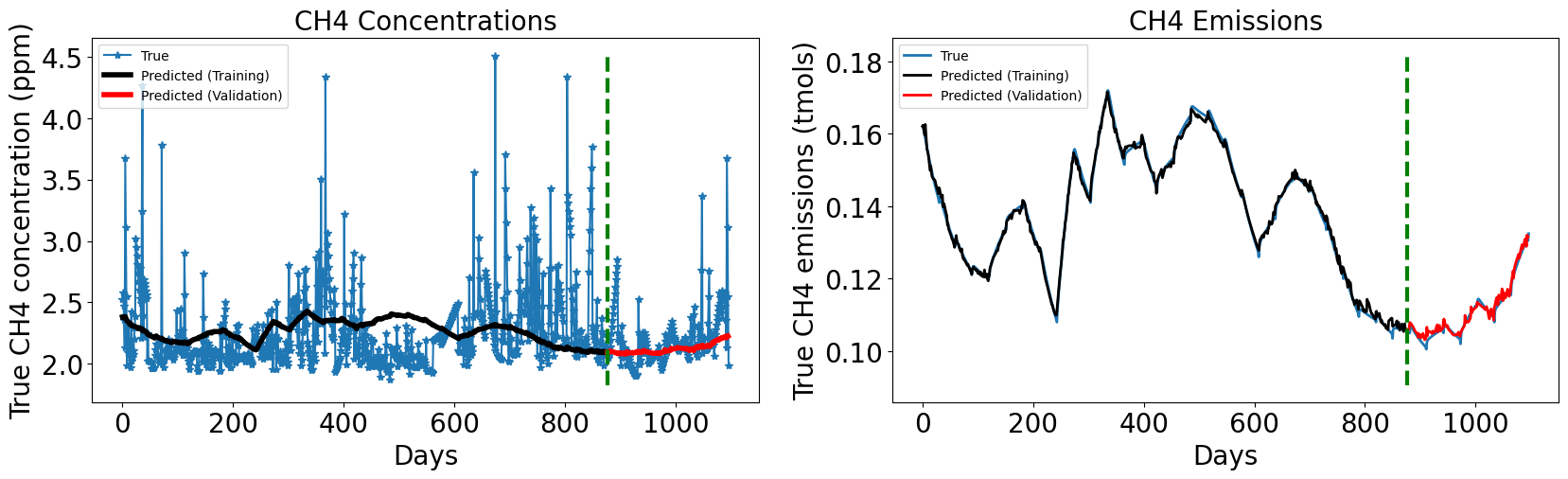}
      \caption{RNN}
     \label{fig:Mannix_oneConnect}
     \end{subfigure}
  \begin{subfigure}
     \centering
      \includegraphics[scale=0.2]{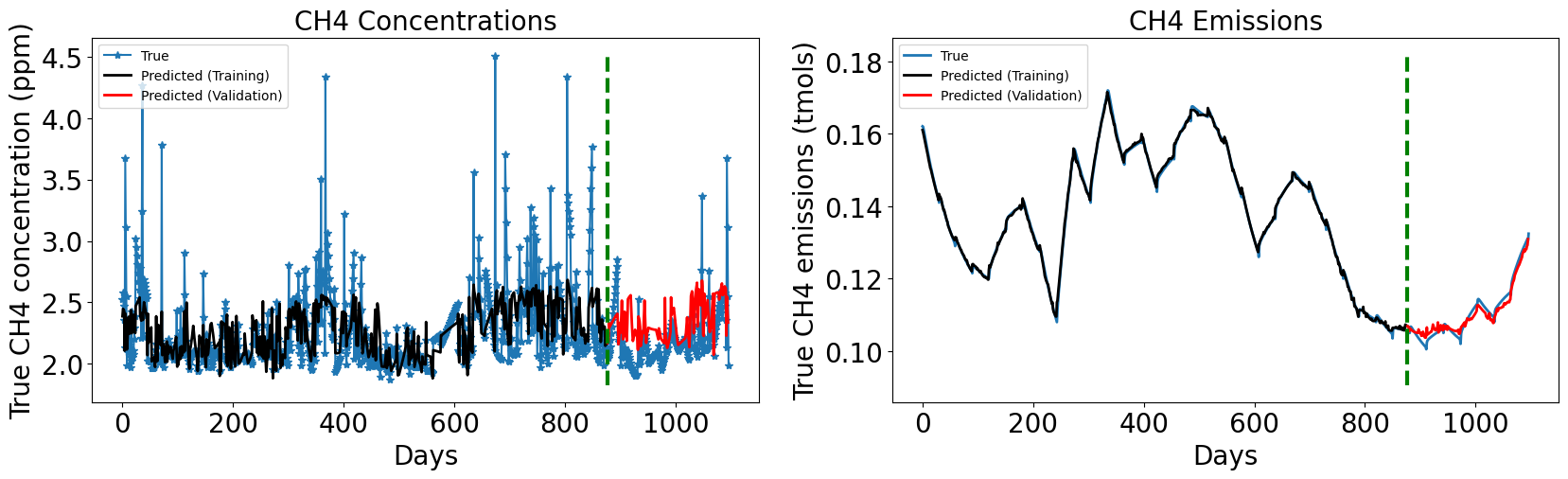}
      \caption{NN}
     \label{fig:Mannix_nn}
 \end{subfigure}
  \begin{subfigure}
     \centering
      \includegraphics[scale=0.2]{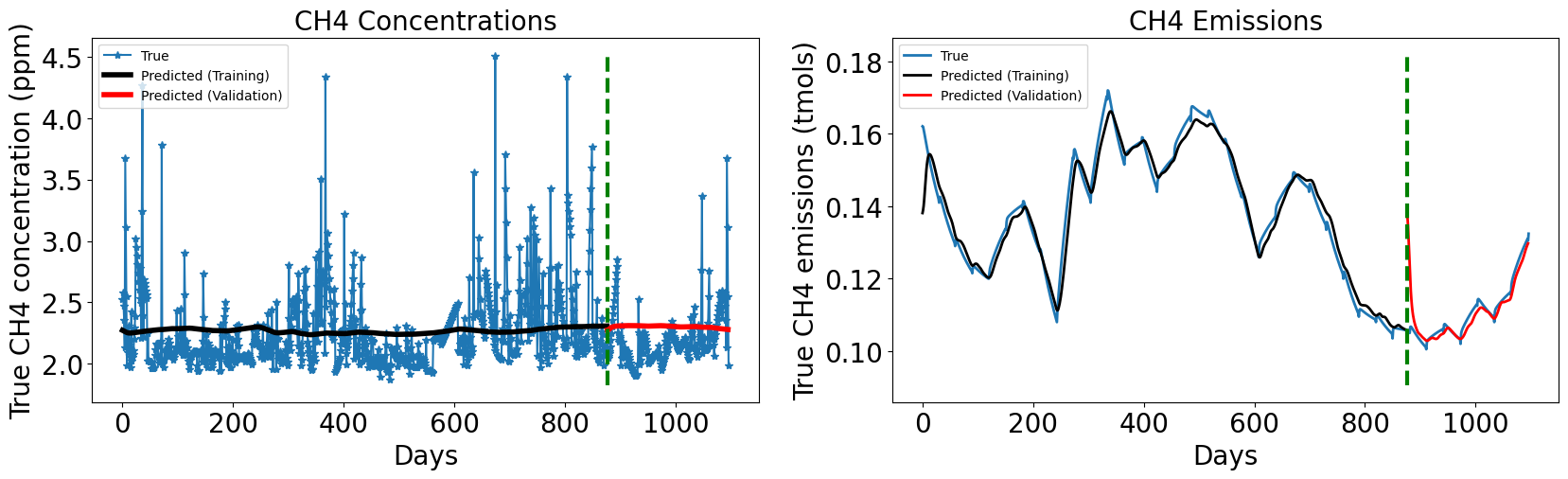}
      \caption{LSTM}
     \label{fig:Mannix_lstm}
 \end{subfigure}

    \caption{Results for weather station Mannix and tailing pond Pond 2/3 for different architectures. The green dashed line indicates the day that splits the data into training and validation set. Before the green dashed line is the training dataset and after the green dashed line is the validation set.}
    \label{fig:Mannix_full}
\end{figure}

The station's nearest active tailing pond is Pond 2/3 owned by Suncor. $\mathbf{x}_{atm}$ for this station was built by considering data from WBEA with wind directions filtered between 280 and 340 degrees. The other part $\mathbf{x}_{dil}$ was built by solving the MMs using diluent data reported by Suncor. We assume about 15\% of the diluents reported by Suncor going into Pond 2/3 based on the FFT volume of this pond. The results for each of the methods is given in Figures \ref{fig:Mannix_full}.
All the variants of the proposed model (forward, reverse and poly) successfully learn the trends of the true concentrations as well as simulated emissions from given MM as depicted in Figures \ref{fig:Mannix_for}, \ref{fig:Mannix_rev} and \ref{fig:Mannix_spapoly}. From Table \ref{tab:rel_errors}, we see that forward or polynomial representations have the lowest error for predicting concentrations outperforming all other models. On the other hand for emissions, since the concentration data is noisy, we also see certain noise/oscillations present in the emission predictions. 
The results from RNN model shows that can learn certain dependencies between the two output quantities. However, excluding $\mathbf{x}_{atm}$ from the constraint leads to the model not learning the oscillations caused by atmospheric parameters in the concentration data. 
For the other methods that are trained without any physical constraints, we see that the full NN model performs comparable to the proposed model (see Figure \ref{fig:Mannix_nn}). 
It learns the trends of methane concentrations and the smooth solution of the MMs. However, this model is equivalent to combining two separate models for learning concentrations and emissions and does not capture dependencies between the two outputs. The LSTM model fails to learn anything from the concentration data, but can easily learn the emission data given that it is simulated from MM as depicted in Figure \ref{fig:Mannix_lstm}. 

\subsubsection{Lower Camp}\label{sec:lc}
\begin{figure}[h!]
    \centering
   \begin{subfigure}
       \centering
      \includegraphics[scale=0.2]{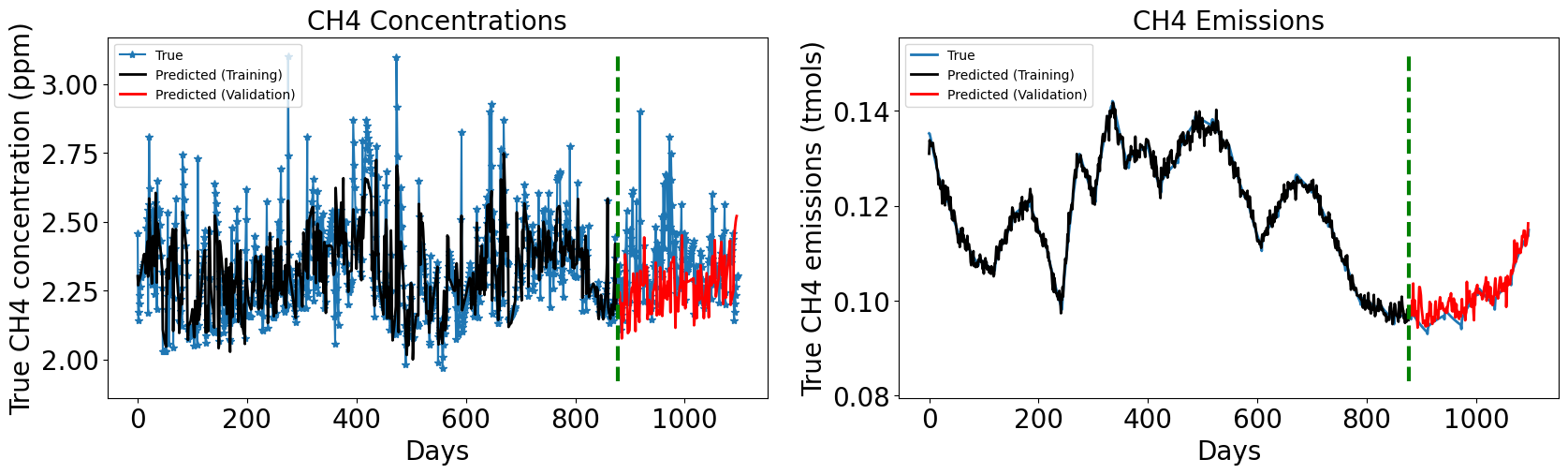}
      \caption{Forward ADM Constraint}
     \label{fig:lc_for}
   \end{subfigure}
   \begin{subfigure}
       \centering
      \includegraphics[scale=0.2]{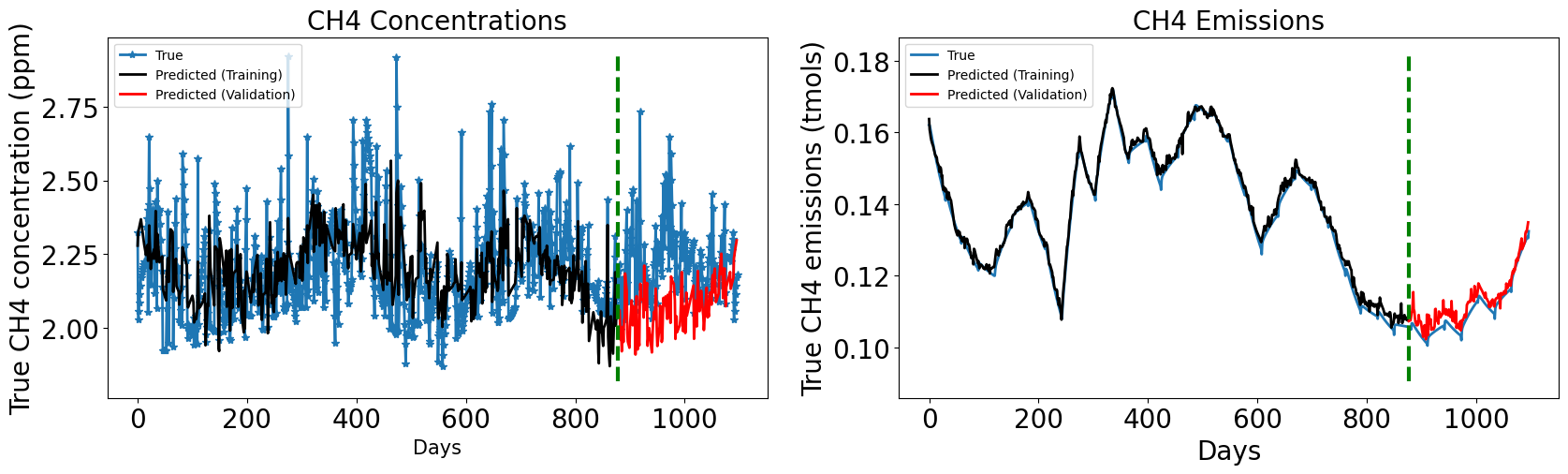}
      \caption{Reverse ADM Constraint}
     \label{fig:lc_rev}
   \end{subfigure}
 \begin{subfigure}
       \centering
      \includegraphics[scale=0.2]{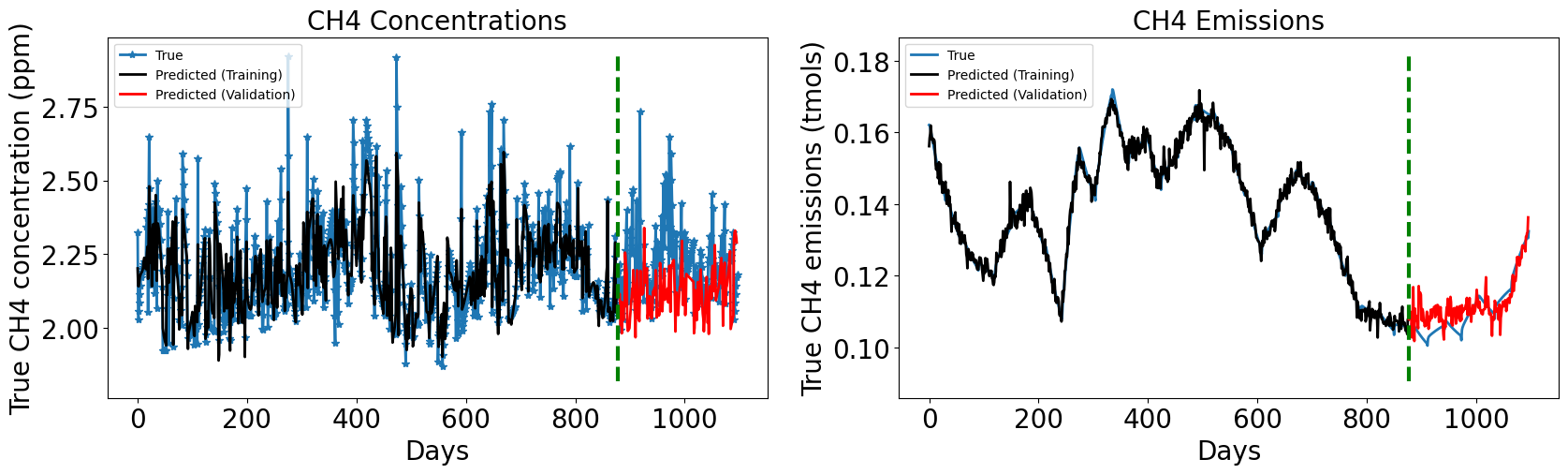}
      \caption{Polynomial Representation of forward ADM constraint}
     \label{fig:lc_spapoly}
   \end{subfigure}
   \begin{subfigure}
       \centering
      \includegraphics[scale=0.2]{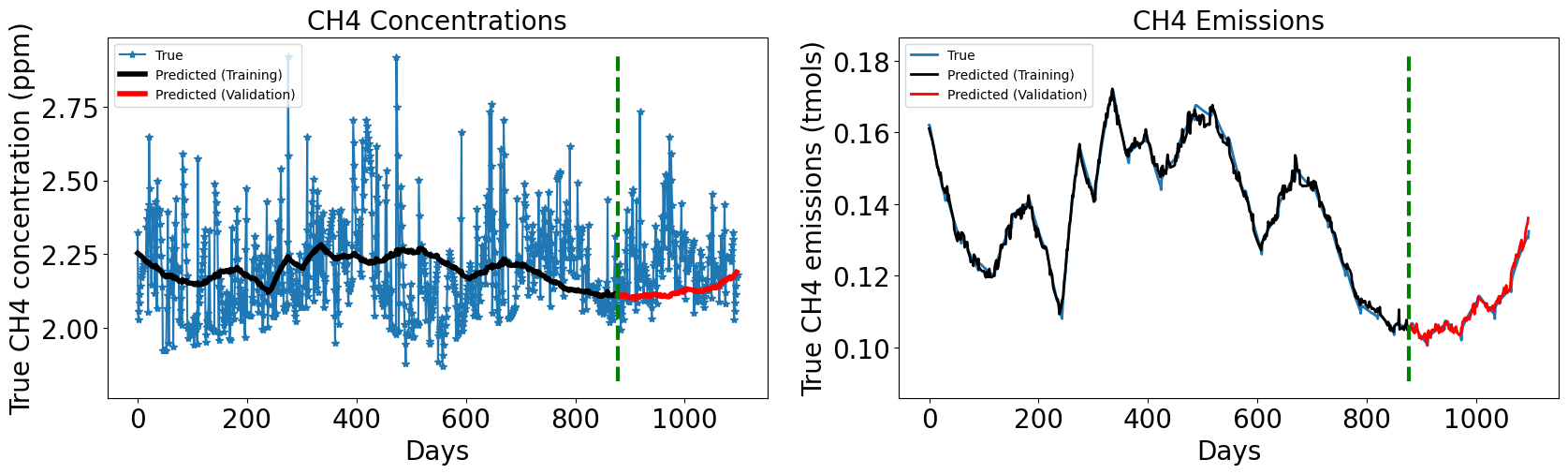}
      \caption{RNN}
     \label{fig:lc_one}
   \end{subfigure}
   
    \begin{subfigure}
       \centering
      \includegraphics[scale=0.2]{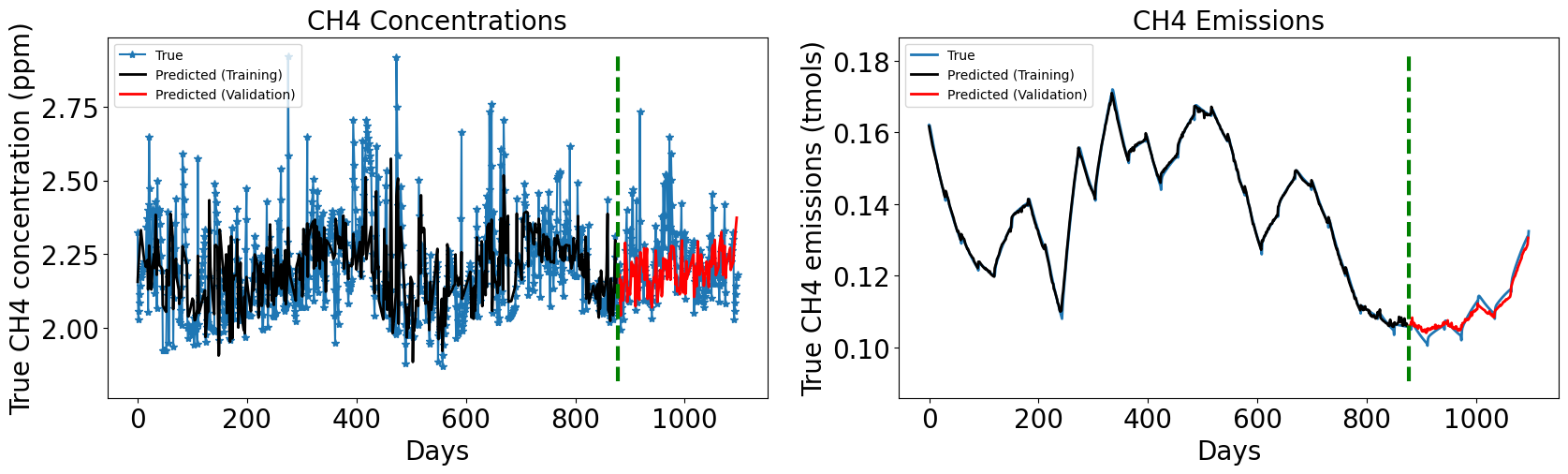}
      \caption{Two layer NN}
     \label{fig:lc_nn}
   \end{subfigure}
   \begin{subfigure}
       \centering
      \includegraphics[scale=0.2]{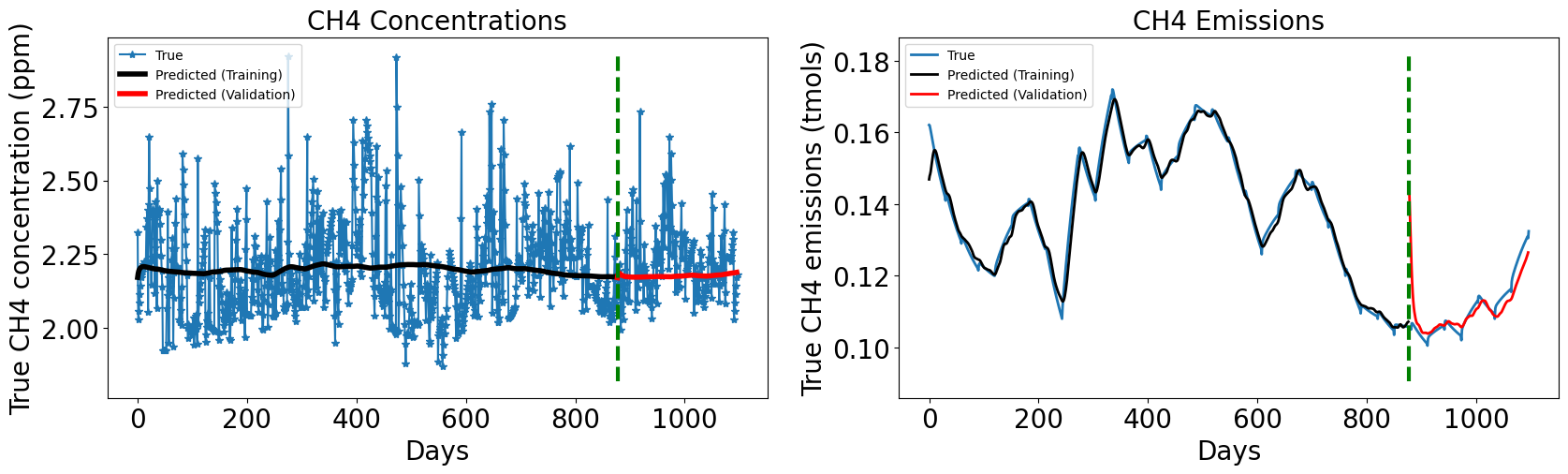}
      \caption{LSTM}
     \label{fig:lc__lstm}
   \end{subfigure}
    \caption{Results for weather station Lower Camp and tailing pond Pond 2/3 for different architectures. The green dashed line indicates the day that splits the data into training and validation set. Before the green dashed line is the training dataset and after the green dashed line is the validation set.}
    \label{fig:lc_full}
\end{figure}

Out of the two OSTPs (Pond 2/3 and Pond 5) near Lower Camp, only Pond 2/3 is active. Thus, we we omit experiments on Pond 5 and build training dataset for Pond 2/3 only. 
We assume about 15\% of the diluents reported by Suncor going to Pond 2/3 based on the FFT volume of this pond.
As in previous cases, the our models and NN perform the best for concentration predictions For MM based emission simulations, while our models learn the trend of emissions with some oscillations directly connected to concentration trends. 
The RNN model fails to learn the concentration data due to absence of any atmospheric parameters in defining the constraint and the LSTM model just learns a straight line that through the concentration dataset (see Figures \ref{fig:lc_one} and \ref{fig:lc__lstm}). For learning the emission data, all the models learn it well since it is simulated from a known MM. 

\subsubsection{Mildred Lake}
\begin{figure}[h!]
    \centering
    \begin{subfigure}
     \centering
      \includegraphics[scale=0.2]{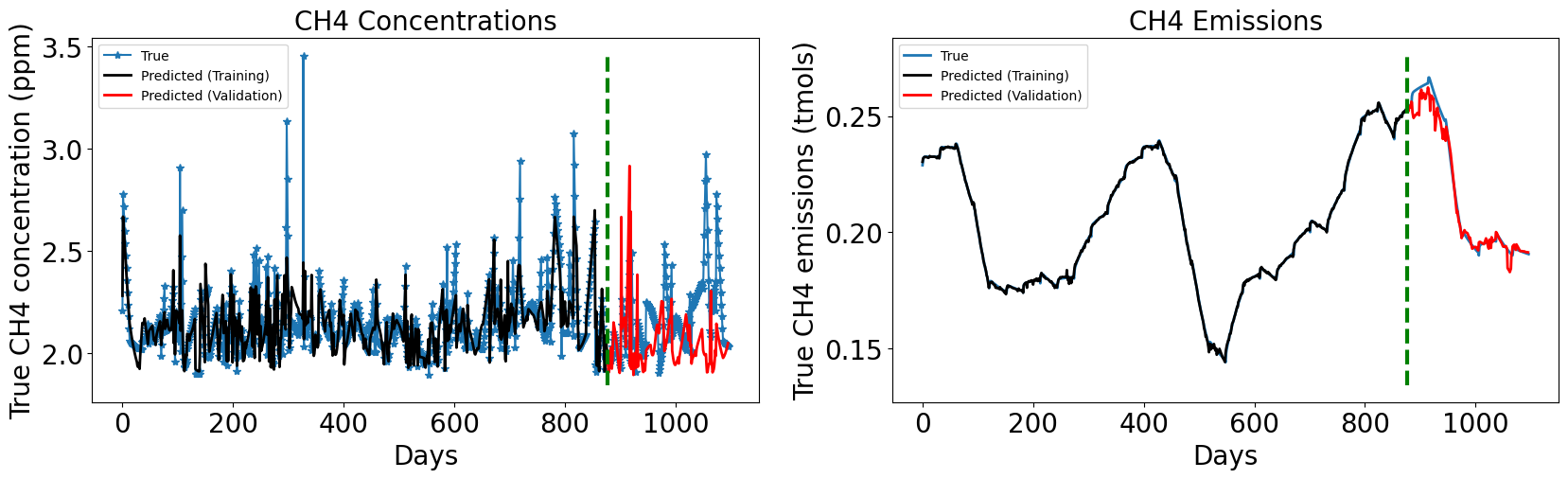}
      \caption{Forward ADM constraint}
     \label{fig:mlsb_for}
 \end{subfigure}
  \begin{subfigure}
     \centering
      \includegraphics[scale=0.2]{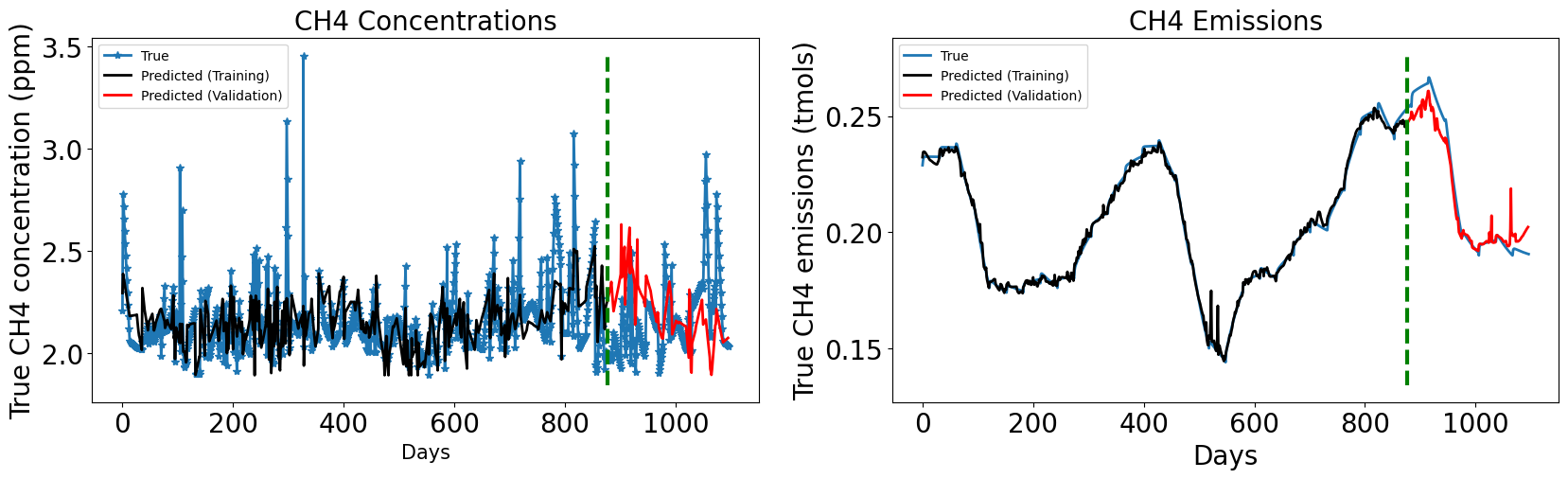}
      \caption{Reverse ADM constraint}
     \label{fig:mlsb_rev}
 \end{subfigure}
 \begin{subfigure}
     \centering
      \includegraphics[scale=0.2]{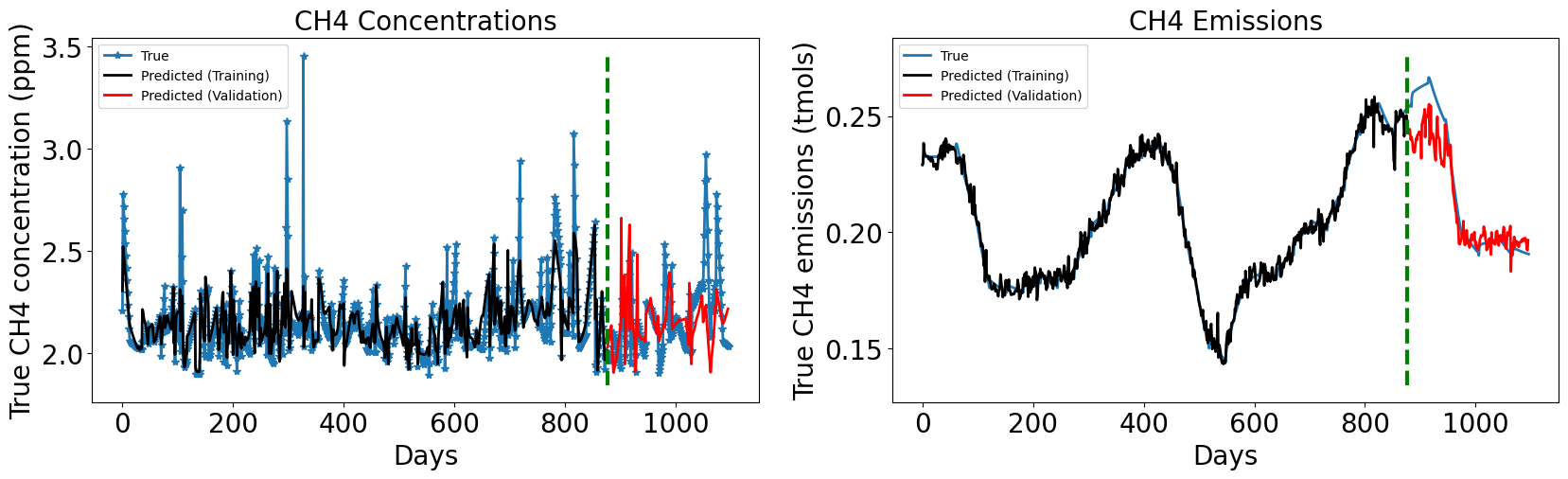}
      \caption{Polynomial Representation of forward ADM constraint}
     \label{fig:mlsb_spapoly}
 \end{subfigure}
 \begin{subfigure}
     \centering
      \includegraphics[scale=0.2]{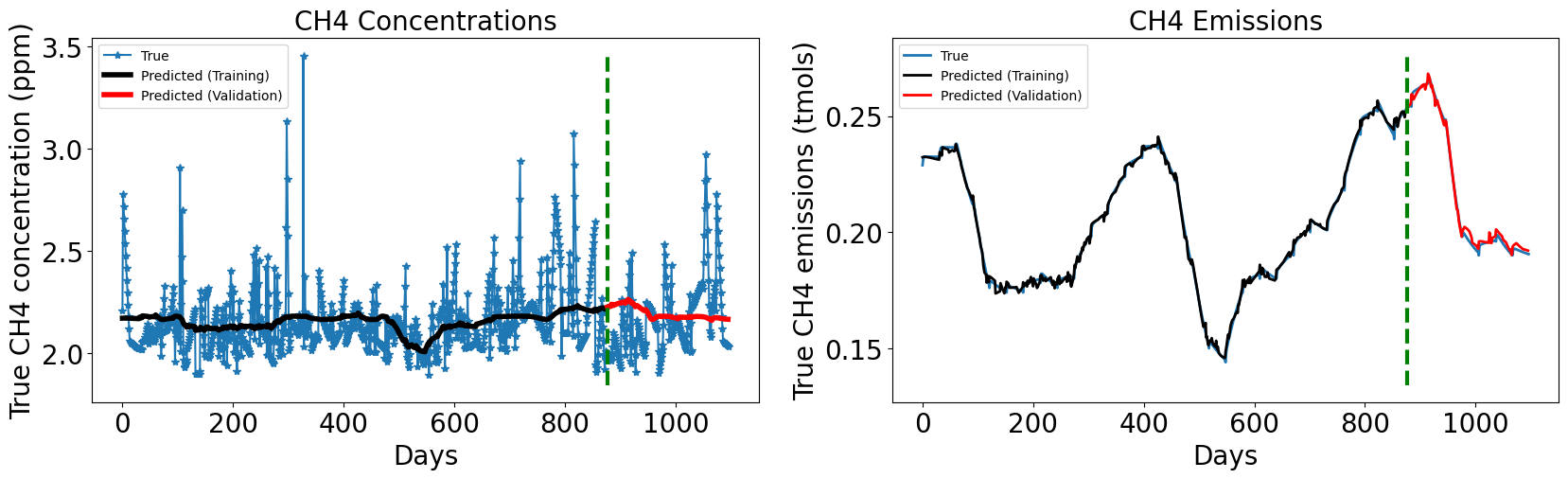}
      \caption{RNN}
     \label{fig:mlsb_one}
 \end{subfigure}
 \begin{subfigure}
     \centering
      \includegraphics[scale=0.2]{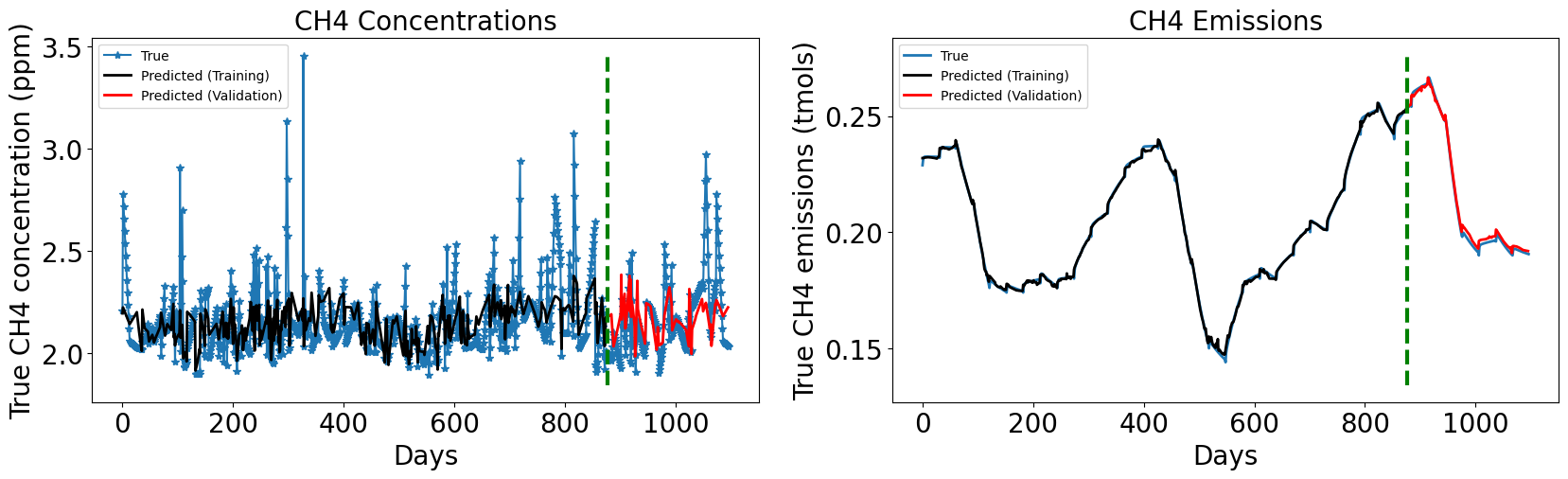}
      \caption{NN}
     \label{fig:mlsb_nn}
 \end{subfigure}
 \begin{subfigure}
     \centering
      \includegraphics[scale=0.2]{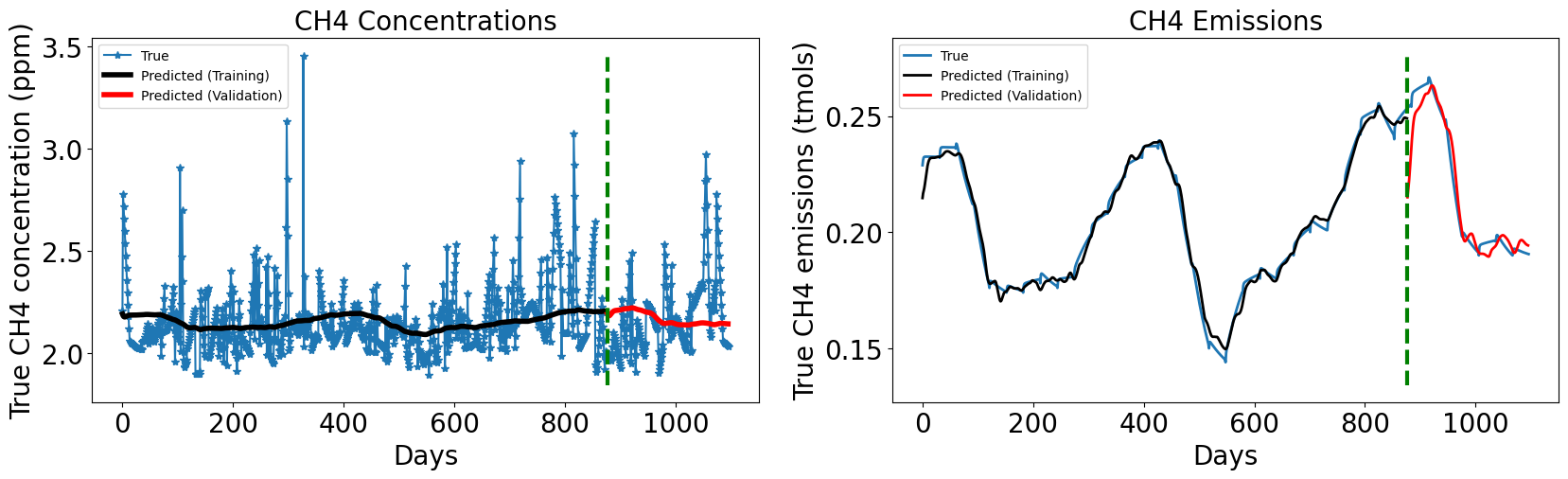
      }
      \caption{LSTM}
     \label{fig:mlsb_lstm}
 \end{subfigure}
        \caption{Results for weather station Mildred Lake and tailing pond Mildred Lake Settling Basin (MLSB) for different architectures. The green dashed line indicates the day that splits the data into training and validation set. Before the green dashed line is the training dataset and after the green dashed line is the validation set.}
    \label{fig:mlsb_full}
\end{figure}

Station Mildred Lake which is located near the Mildred Lake Settling Basin (MLSB) which is an active tailing pond owned by Syncrude.
Another significant range of the wind direction for this station corresponds to Pond 5 (Suncor) which is inactive (see Figure \ref{fig:study_area}) and thus excluded from model training. 
Further, we assume about 20\% of the diluents reported by Syncrude going to Pond MLSB based on the FFT volume of this pond.
From Figure \ref{fig:mlsb_full}, we see the our models best fit the concentration data with lowest error given by polynomial fitted model (see Table \ref{tab:rel_errors}). This is closely followed by the NN model. Comparing Figures \ref{fig:mlsb_for}, \ref{fig:mlsb_rev}, \ref{fig:mlsb_spapoly} and \ref{fig:mlsb_nn} we see that the our trained models are much better at fitting well with majority of the peaks on concentration data (except the extremely sharp one beyond 2.5 ppm as they are considered as outliers in the dataset) than the NN model. 
The RNN model can only pickup the trend of concentration data while fitting almost perfectly with the emission data.
The LSTM model simply fails to generalize anything in the concentration data. For emission data, the trends are similar to the previous stations where the unconstrained formulations (NN and LSTM) learn the smooth function almost perfectly.
\subsection{Tracking Emissions around Stations}
\begin{figure}[h!]
    \centering
    \includegraphics[scale=0.5]{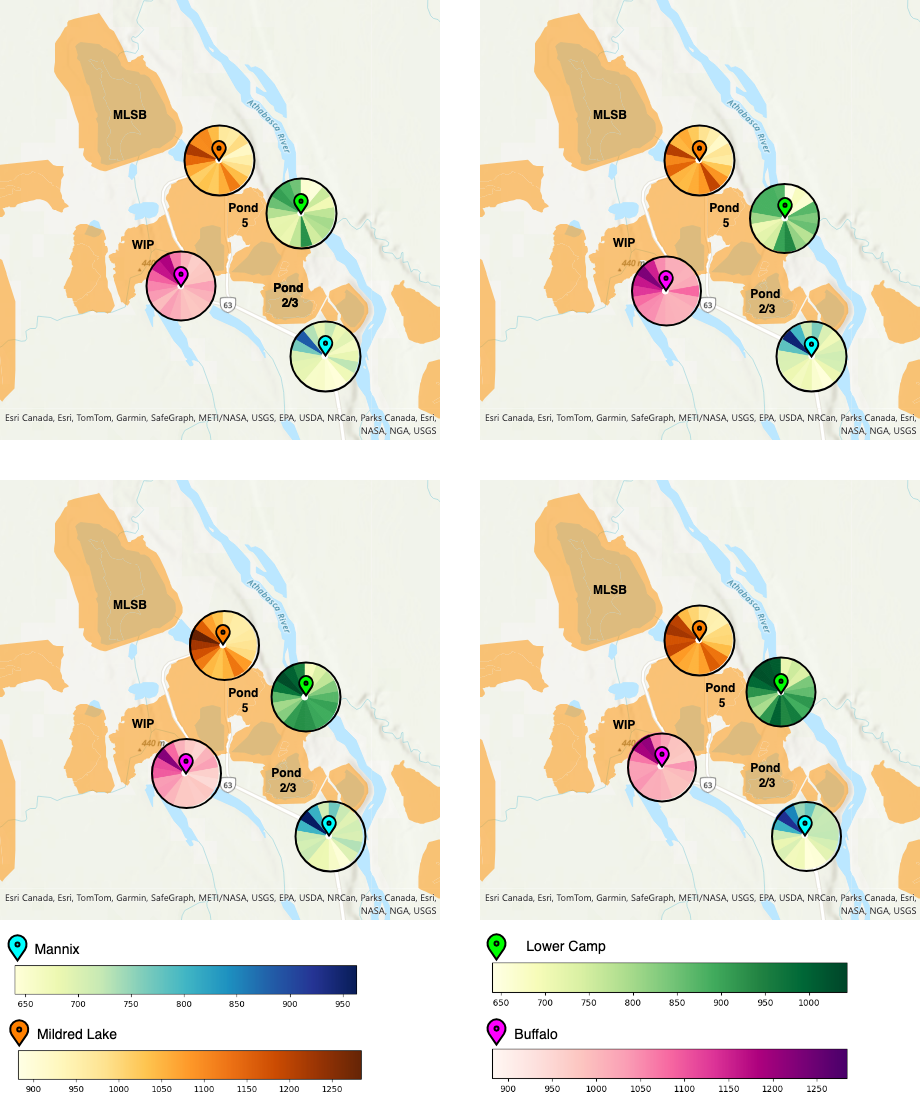}
        \caption{Emissions (in tonnes) per year predicted by DIRNN from around the weather stations in our region of case study over four years. \textbf{Top left:} Year 2020. \textbf{Top right:} Year 2021, \textbf{Bottom left:} Year 2022. \textbf{Bottom right:} Year 2023.}
    \label{fig:emm_track}
\end{figure}

Plots for all the years between 2020 and 2023 in Figure \ref{fig:emm_track}. See main text for full details.
We also use the method WindTrax to compare emission estimates for Pond 2/3 when considered concentration measurements from station Lower Camp with wind directions from 160-180 degrees. However, as we are working with limited data, the outputs given by the WindTrax software for years 2020-2023 give unrealistic estimates of more than $10^5$ t/y. The parameters used for generating the estimates are given below.
\begin{enumerate}
    \item Area of source: 2.8 $km^2$.
\item Distance of measuring station to source: 499 m.
\item Location of monitoring station: North of the source.
4. Height of measuring station: 10 m.
\end{enumerate}
The outputs can be seen in separately submitted 'Output Data WindTrax' file in the submission.

\subsection{Estimating Total Emissions}
Since data for all the ponds are not available and neither are there weather stations around them, it is not possible to train a model to get individual emission levels. Thus, using a sample of few OSTPs and weather stations, we train DIRNN and use those results to estimate the emission levels for all the OSTPs in the Athabasca region. Since the FFT volume is directly connected to level of microbial activity (and thus CH$_4$ emissions), we used the total FFT volume of all the OSTPs and EPLs to compute our target. The detailed computation and results are given below. First, we apply the reverse formulation of our trained model to obtain emission estimates for each weather station for every 20 degree interval. Selecting the directions towards the OSTPs/EPLs, we add the estimates from the respective direction intervals to compute the CH$_4$ emission levels from each pond.

\begin{table}[h!]
    \centering
    \begin{tabular}{|c|c|c|c|c|c|}
    \hline
         &  MLSB & Pond 2/3 & Pond 5 & WIP & Total\\
         \hline
        FFT Volume ($Mm^3$) & 110.5 & 36.9 & 30.4 & 166.8 & 344.6\\
        Model estimation (t) & 4790.94& 2895.3 & 1913 & 2398 & 11997.24\\
        \hline
    \end{tabular}
    \caption{Estimated emissions from the OSTPs in the dataset for 2023.}
    \label{tab:my_label}
\end{table}
The total FFT volume of all the companies in the Athabasca region is 1617.2 $Mm^3$. Thus using the numbers from the table above, we can compute each 1 $Mm^3$ of FFT contributing to 34.81 t of CH$_4$ per year, leading to a total of 56,302.77 t of CH$_4$ from all OSTPs combined. This is almost 3.5 times the reported number of 15736 t in \citep{nir}.
\subsection{Emissions Simulated with a Different Methanogenesis Model}\label{sec:2gen}

\begin{figure}[h!]
    \centering
    \begin{subfigure}
     \centering
      \includegraphics[scale=0.2]{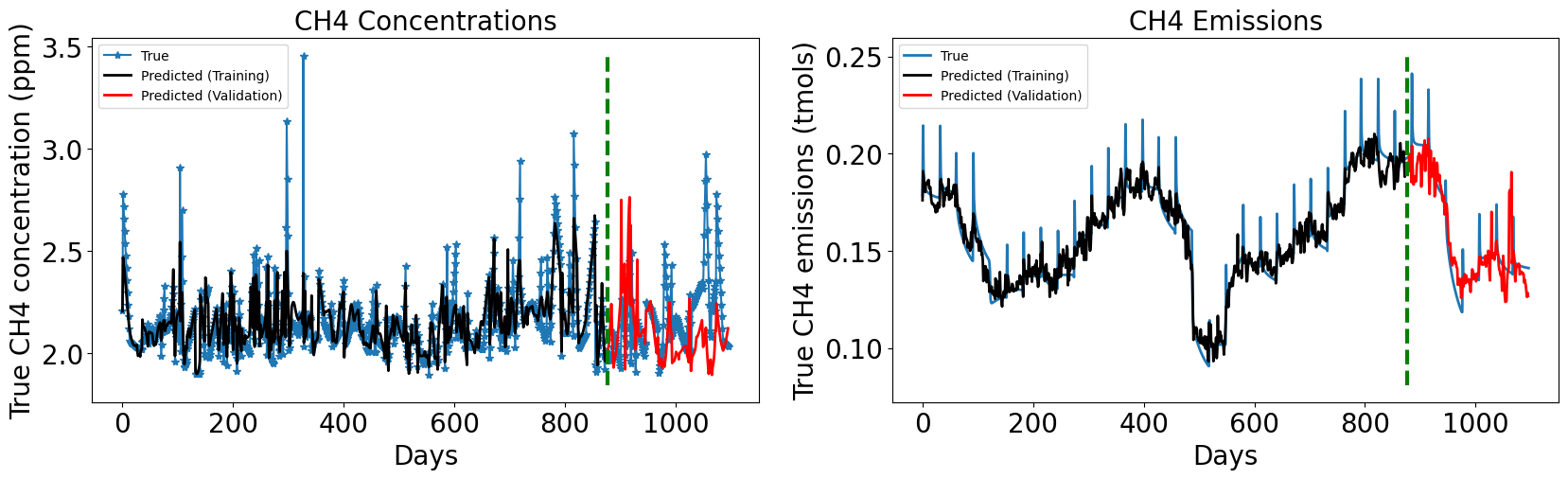}
      \caption{Forward ADM constraint}
     \label{fig:secgen_for}
 \end{subfigure}
 \begin{subfigure}
     \centering
      \includegraphics[scale=0.2]{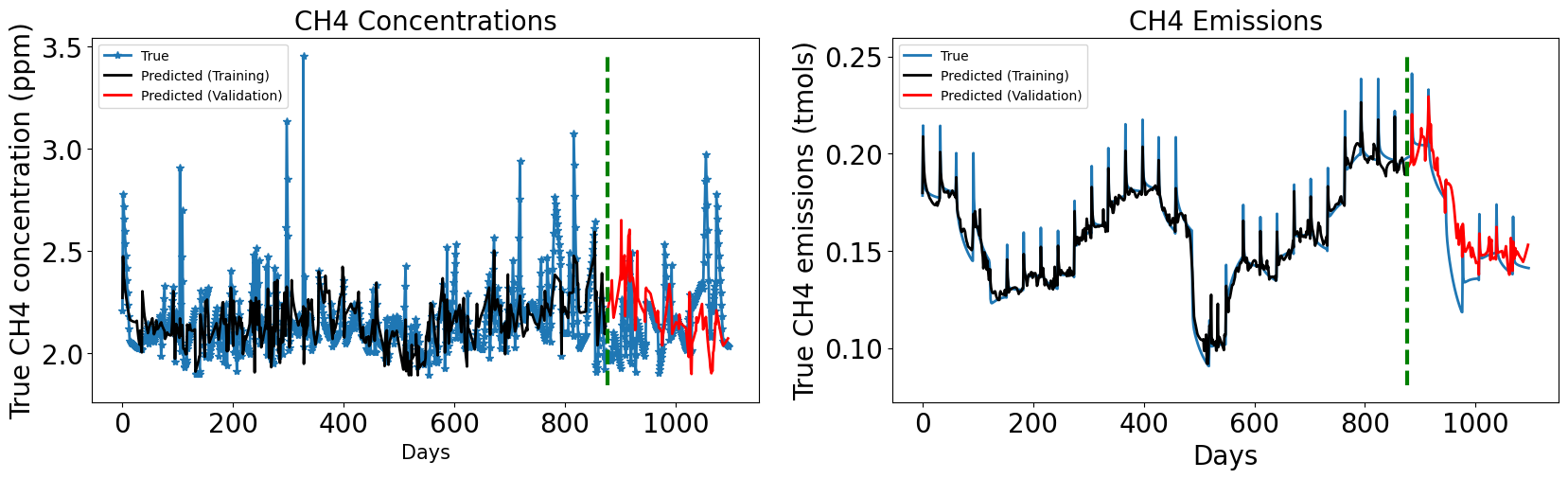}
      \caption{Reverse ADM constraint}
     \label{fig:secgen_rev}
 \end{subfigure}
 \begin{subfigure}
     \centering
      \includegraphics[scale=0.2]{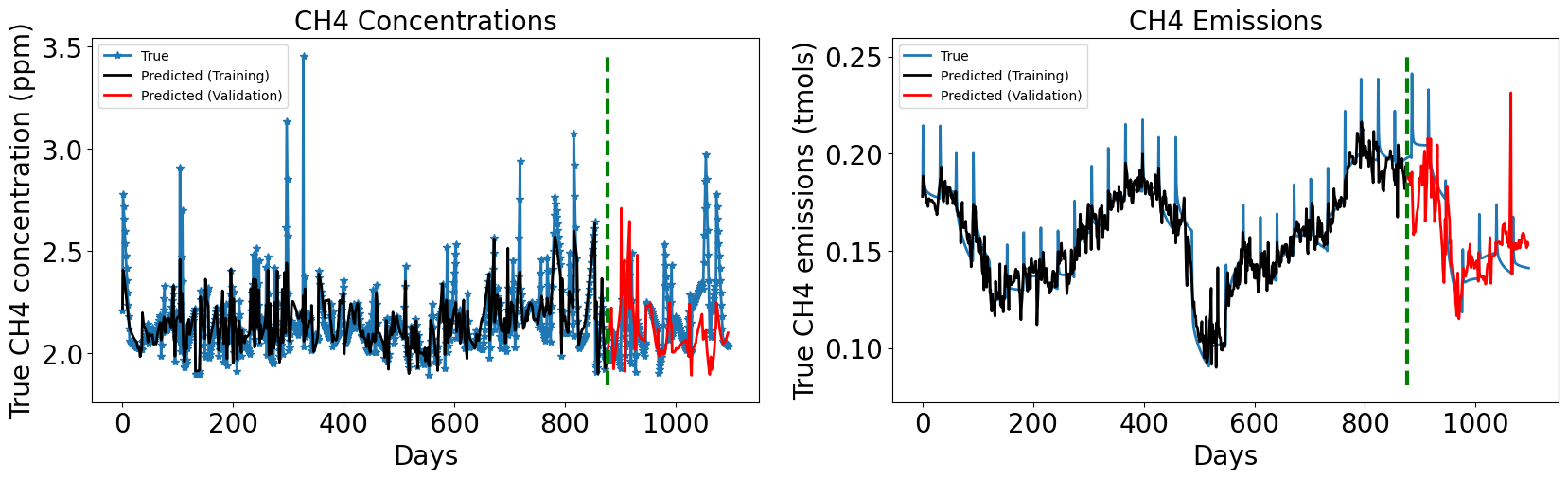}
      \caption{Polynomial Representation of forward ADM constraints}
     \label{fig:secgen_spapoly}
 \end{subfigure}
  \begin{subfigure}
     \centering
      \includegraphics[scale=0.2]{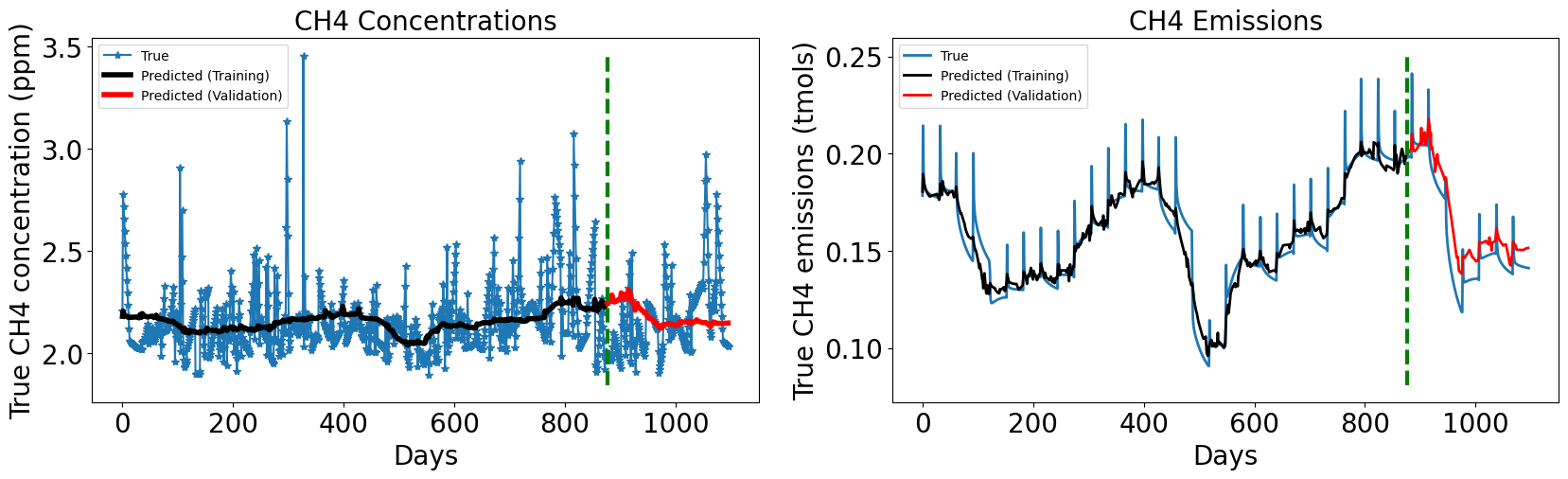}
      \caption{RNN}
     \label{fig:secgen_oneconnect}
 \end{subfigure}
  \begin{subfigure}
     \centering
      \includegraphics[scale=0.2]{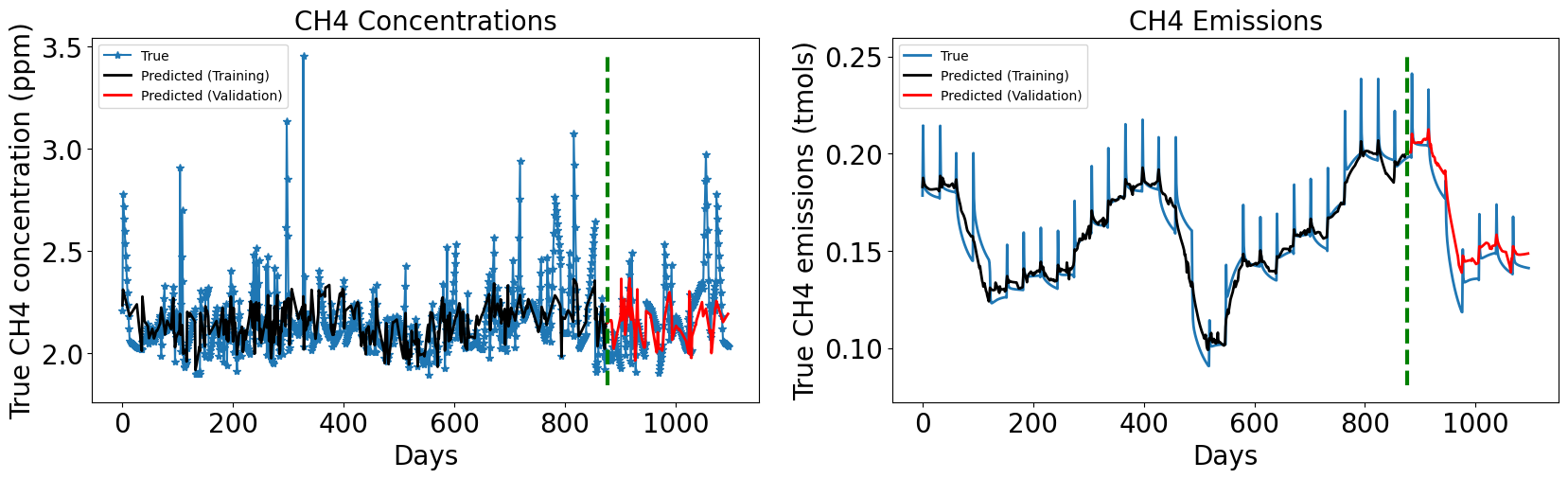}
      \caption{NN}
     \label{fig:secgen_nn}
 \end{subfigure}
  \begin{subfigure}
     \centering
      \includegraphics[scale=0.2]{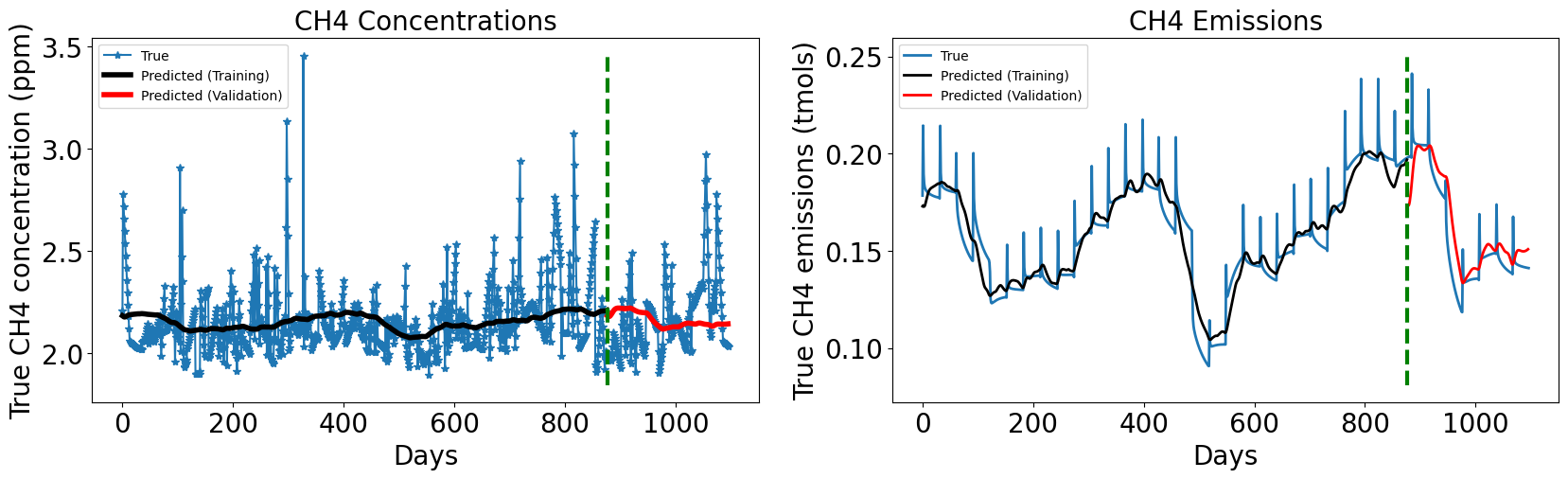}
      \caption{LSTM}
     \label{fig:secgen_lstm}
 \end{subfigure}
    \caption{Results for Mildred Lake for MLSB pond for different architectures using a different MM from \citep{kong2019second}.}
    \label{fig:second_gen}
\end{figure}

In this paper we aim to train a model than can fill gaps between MMs estimating methane emissions and the atmospheric methane concentrations measured by weather monitoring stations. Our previously generated results were based on using a temperature dependent MM. In this section we show that this approach can be used with any methanogenesis model.
We trained our model(s) using data generated from the second generation stoichiometric model \citep{kong2019second}.
The plots for predicting concentrations and emissions are given in Figure \ref{fig:second_gen}. As with previous simulations, we see that our models outperform all the other models for learning the concentration data. 
Similar conclusions can be made for the results using this model as well.

\pagebreak

\begin{table}
    \centering
    \caption{List of abbreviations used in the paper}
    \begin{tabular}{|c|c|}
    \hline
        Abbreviation & Description \\
        \hline
        CH$_4$ & Methane\\
        CO$_2$ & Carbon di Oxide \\
        OSTP & Oil Sands Tailing Pond \\
        EPL & End Pit Lake \\
        FFT & Fine Fluid Tailings \\
        ML & Machine Learning \\
        MM & Mechanistic Model\\
        RE & Relative Error\\
        NN & Neural Network \\
        LSTM & Long Short Term Memory \\
        RNN & Recurrent Neural Network \\
        PDE & Partial Differential Equation\\
        WBEA & Wood Buffalo Environment Association\\
        t & Tonnes \\
        MtCO$_2$e & Million tonnes of carbon dioxide equivalent \\
        ppm & Parts per million \\ 
        ppb & Parts per billion\\
        std & standard deviation\\
        kmph & kilometer per hour\\
        \hline
    \end{tabular}
    \label{tab:abbr}
\end{table}

\begin{table}[h!]
    \centering
    \begin{tabular}{|c|c|c|}
    \hline
        Variable & Type & Units \\
        \hline
        	Atm. Temperature & $\mathbf{x}_{atm}$ & Celcius \\
            Hydrogen Sulphide &  $\mathbf{x}_{atm}$ & ppb\\
            Relative humidity & $\mathbf{x}_{atm}$ & Percentage \\
            Sulphur di oxide & $\mathbf{x}_{atm}$ & ppb \\
            Wind direction & $\mathbf{x}_{atm}$ & Degrees \\
            Wind direction std &$\mathbf{x}_{atm}$& Degrees\\
            Wind speed & $\mathbf{x}_{atm}$ & kmph \\
            Wind speed std & $\mathbf{x}_{atm}$ & kmph\\
            Barometric pressure & $\mathbf{x}_{atm}$ & milibar \\
            Dew point & $\mathbf{x}_{atm}$ & Celcius \\
            Solar radiation & $\mathbf{x}_{atm}$ & Watts per square meter \\
            Vertical wind & $\mathbf{x}_{atm}$ & kmph\\
            Vertical wind std & $\mathbf{x}_{atm}$ & kmph \\
            \hline
            
    \end{tabular}
    \caption{List of atmospheric variables used as inputs. \textcolor{black}{All the data forr these input variables are collected by the weather monitoring stations under the WBEA. Since the study area is chosen based on the location of the weather monitoring stations, data for all the variables is available at each of the site under consideration. Data is collected hourly.}}
    \label{tab:var}
\end{table}

\end{document}